\documentclass[preprint,showpacs,preprintnumbers,amsmath,amssymb,superscriptaddress]{revtex4}

\usepackage{graphicx}
\usepackage{dcolumn}
\usepackage{bm}

\font\scripti=cmmi7
\font\scriptscripti=cmmi5
\def\sib#1{\setbox0 = \hbox{\scripti #1}
  \kern-.02em\copy0\kern-\wd0
  \kern.04em\box0} 
\def\ssib#1{\setbox0 = \hbox{\scriptscripti #1}
  \kern-.02em\copy0\kern-\wd0
  \kern.04em\box0} 
\font\tenib=cmmib10 
\skewchar\tenib='177 \skewchar\tenib='177 \skewchar\tenib='177
\def\pbold#1{\setbox0 = \hbox{$ #1 $}
  \kern-.022em\copy0\kern-\wd0
  \kern.011em\copy0\kern-\wd0
  \kern.011em\copy0\kern-\wd0
  \kern.011em\copy0\kern-\wd0
  \kern.011em\box0} 

\usepackage{graphicx}
\usepackage{dcolumn}
\usepackage{bm}

\def\lesssim{\ \raise.3ex\hbox{$<$}\kern-0.8em\lower.7ex\hbox{$\sim$}\ }
\def\gesim{\ \raise.3ex\hbox{$>$}\kern-0.8em\lower.7ex\hbox{$\sim$}\ }

\def\dag{\dagger}

\def\sig{\sigma}
\def\Sig{\Sigma}
\def\om{\omega}
\def\Gam{\Gamma}

\begin{document}
\title{Excitation properties and effects of mass imbalance in the BCS-BEC crossover regime of an ultracold Fermi gas}

\author{Ryo Hanai}
\email{rhanai@rk.phys.keio.ac.jp}
\affiliation{Department of Physics, Keio University, 3-14-1 Hiyoshi, Kohoku-ku, Yokohama 223-8522, Japan} 

\author{Takashi Kashimura}
\affiliation{Department of Physics, Keio University, 3-14-1 Hiyoshi, Kohoku-ku, Yokohama 223-8522, Japan} 

\author{Ryota Watanabe}
\affiliation{Department of Physics, Keio University, 3-14-1 Hiyoshi, Kohoku-ku, Yokohama 223-8522, Japan} 

\author{Daisuke Inotani}
\affiliation{Department of Physics, Keio University, 3-14-1 Hiyoshi, Kohoku-ku, Yokohama 223-8522, Japan} 
\affiliation{Graduate School of Pure and Applied Sciences, University of Tsukuba, Tsukuba, Ibaraki 305-8571, Japan}

\author{Yoji Ohashi}
\affiliation{Department of Physics, Keio University, 3-14-1 Hiyoshi, Kohoku-ku, Yokohama 223-8522, Japan} 

\date{\today}
\begin{abstract}
We investigate single-particle properties of a mass-imbalanced Fermi gas in the BCS (Bardeen-Cooper-Schrieffer)-BEC (Bose-Einstein condensation) crossover region. In the presence of mass imbalance, we point out that the ordinary $T$-matrix approximation, which has been extensively used to clarify various BCS-BEC crossover physics in the mass-balanced case, unphysically gives a double-valued solution in terms of the superfluid phase transition temperature $T_{\rm c}$ in the crossover region. To overcome this serious problem, we include higher order strong-coupling corrections beyond the $T$-matrix level. Using this extended $T$-matrix theory, we calculate single-particle excitations in the normal state above $T_{\rm c}$. The so-called pseudogap phenomena originating from pairing fluctuations are shown to be different between the light mass component and heavy mass component, which becomes more remarkable at higher temperatures. Since Fermi condensates with hetero-Cooper pairs have recently been discussed in various fields, such as exciton (polariton) condensates, as well as color superconductivity, our results would be useful for the further development of Fermi superfluid physics, beyond the conventional superfluid state with homo-Cooper pairs.
\end{abstract}

\maketitle
\par
\section{Introduction}
\par
Since the achievement of the superfluid phase transition in $^{40}$K\cite{Regal2004} and $^6$Li\cite{Zwierlein,Kinast,Bartenstein} Fermi gases, various possibilities beyond this simplest $s$-wave pairing state have been discussed in this field, such as a $p$-wave superfluid\cite{Regal,Zhang,Schunck,Ohashi,Ho,Gurarie,Levinsen,Iskin,Inotani}, the Sarma phase\cite{Sarma1963,Liu2003,Forbes2005,Sheehy2007}, a dipolar Fermi condensate\cite{Baranov2002,Bruun2008,Ni2010,Wu2012,Heo2012}, and a mass-imbalanced Fermi superfluid\cite{Wille2008a,Taglieber2008,Voigt2009,Costa2010,Naik2010,Tiecke2010,Lin,Wu2006,Iskin2006b,Iskin2007,Pao,Parish2007,Orso,Baranov,Randeria,Stoof1,Stoof2,Koga}. Although none of these states has been observed yet, the high tunability of cold Fermi gases makes us expect that some of them may be realized in the near future. Once an unconventional pairing state is obtained, one can study its superfluid properties in a wide parameter region, by adjusting various tunable parameters, such as the strength of a pairing interaction\cite{Chin2010}.  Since an ultracold Fermi gas is expected as a useful quantum simulator for the study of strongly correlated fermion systems, an unconventional superfluid Fermi gas would also contribute to the further development of superfluid physics.
\par
Among various possibilities discussed in ultracold Fermi gases, in this paper, we pick up a hetero-pairing state in the presence of mass imbalance\cite{Wille2008a,Taglieber2008,Voigt2009,Costa2010,Naik2010,Spiegelhalder2010,Tiecke2010,Lin,Iskin2007,Pao,Orso,Baranov,Randeria,Stoof1,Stoof2,Koga}. In this state, Cooper pairs are formed between different species with different masses, which is quite different from the ordinary {\it homo}-Cooper pairs in metallic superconductivity. This unique pairing state is expected to realize the Sarma phase\cite{Stoof1}, where single-particle excitations are gapless in spite of a nodeless $s$-wave superfluid order parameter\cite{Sarma1963,Liu2003,Sheehy2007}. In addition, hetero-Cooper pairs have also been discussed in various fields, such as an exciton condensate in a semiconductor\cite{Yoshioka2011,High2012}, exciton-polariton condensate in a semiconductor microcavity\cite{Imamog1996,Tassone1999,Deng2002,Kasprzak2006}, as well as color superconductivity in a dense quark matter\cite{Barrois1977,Bailin1984}. Thus, the mass-imbalanced superfluid Fermi gas is related to various aspects of Fermi superfluid physics. At present, gas mixtures of two different fermionic species\cite{Wille2008a,Taglieber2008,Voigt2009,Costa2010,Naik2010,Tiecke2010,Spiegelhalder2010}, as well as $^{40}$K-$^6$Li hetero-pairs\cite{Voigt2009,Spiegelhalder2010}, have been realized. 
\par
In the current stage of research for a mass-imbalanced Fermi gas, the achievement of the superfluid phase transition is one of the most important issues. For this purpose, the so-called pseudogap phenomenon (which has been extensively discussed in the BCS-BEC crossover regime of a mass-balanced Fermi gas\cite{Chen2009,Magierski2009a,Tsuchiya2009b,Tsuchiya2010,Watanabe2010,Hu2010a,Stewart2008,Gaebler2010}) would be useful, because this precursor phenomenon of the superfluid instability enables to see to what extent the system is close to the superfluid phase transition. Thus, in this paper, we examine single-particle excitations in the BCS-BEC crossover regime of a mass-imbalanced Fermi gas, to clarify pseudogap physics in this system.
\par
\begin{figure}
\begin{center}
\includegraphics[width=0.5\linewidth]{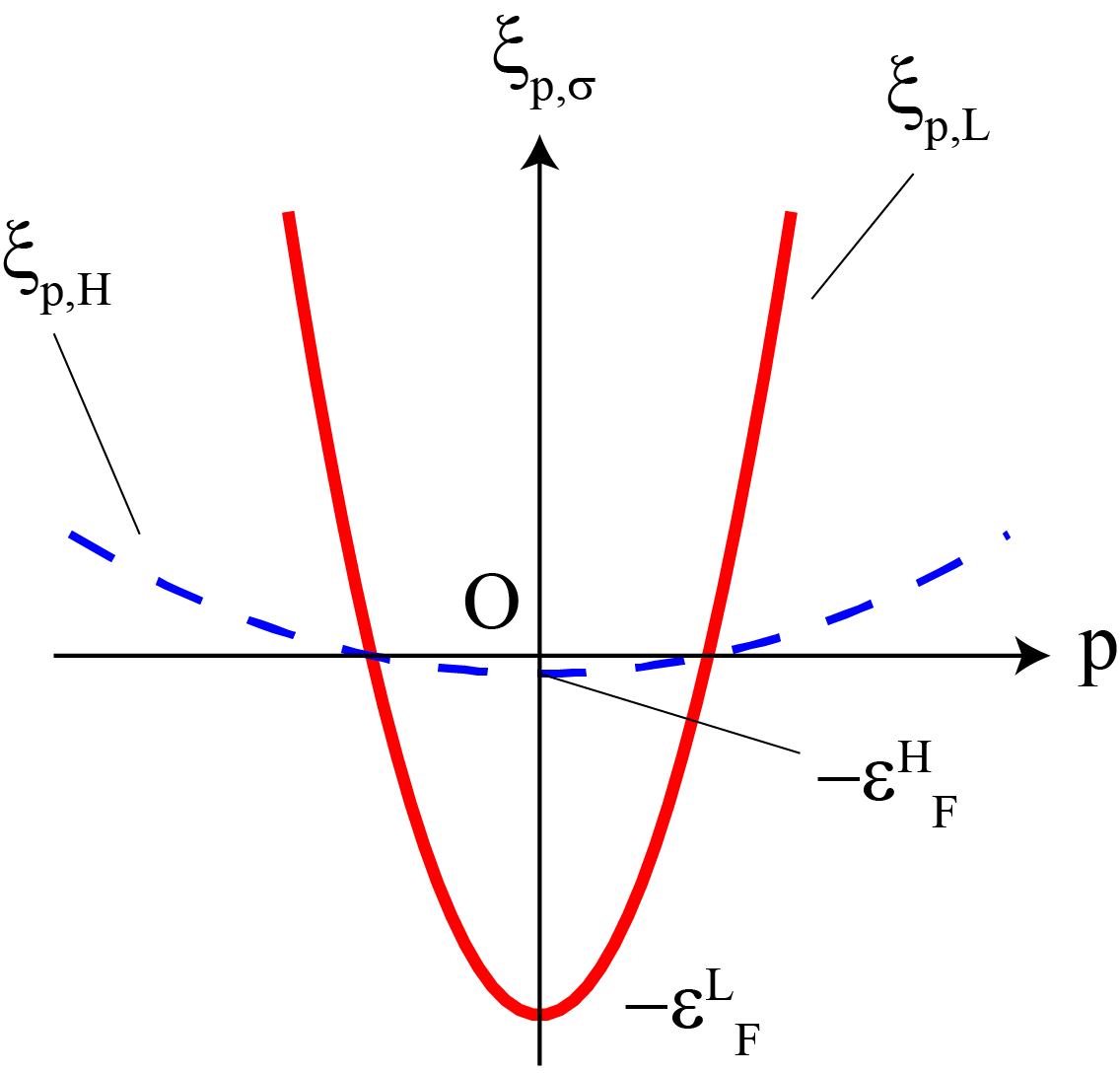}
\end{center}
\caption{(Color online) Schematic single-particle dispersions $\xi_{{\bm p}{\sigma}}=p^2/(2m_\sigma)-\varepsilon_{\rm F}^\sigma$ of a light fermion ($\sigma={\rm L}$) and a heavy fermion ($\sigma={\rm H}$). $\varepsilon_{\rm F}^\sigma$ and $m_\sigma$ are the Fermi energy and an atomic mass of the $\sigma$-component, respectively. In the absence of population imbalance, both the components have the common Fermi momentum, $k_{\rm F}\equiv \sqrt{2m_{\rm L}\varepsilon_{\rm F}^{\rm L}}=\sqrt{2m_{\rm H}\varepsilon_{\rm F}^{\rm H}}$.}
\label{fig1}
\end{figure}
\par
Here, we point out some keys in considering a mass-imbalanced Fermi gas. In the case of a non-interacting gas mixture of $N/2$ light fermions (with mass $m_{\rm L}$) and $N/2$ heavy fermions (with mass $m_{\rm H}$), both the components have the same Fermi surface size at $T=0$, as schematically shown in Fig. \ref{fig1} (where $N$ is the total particle number). Then, since a Cooper pair is usually formed between a light fermion ($\sigma={\rm L}$) at the momentum ${\bm p}$ and a heavy fermion ($\sigma={\rm H}$) at $-{\bm p}$ near the Fermi surface (in the weak-coupling case), the mass imbalance is not so serious for the pair formation. On the other hand, at finite temperatures, the chemical potential $\mu_\sigma$ in the $\sigma$-component behaves as\cite{Mermin}
\begin{equation}
\mu_\sigma(T)=\varepsilon_{\rm F}^\sigma
\Bigl[1-{\pi^2 \over 12}\Bigl({T \over T_{\rm F}^\sigma}\Bigr)^2
\Bigr],
\end{equation}
where $\varepsilon^\sigma_{\rm F}=(3\pi^2N)^{2/3}/(2m_\sigma)$ and $T_{\rm F}^\sigma=\varepsilon^\sigma_{\rm F}/k_{\rm B}$ are the Fermi energy and the Fermi temperature in the $\sigma$-component, respectively. The resulting effective Fermi momenta defined by 
\begin{equation}
{\tilde k}_{{\rm F}\sigma}\equiv\sqrt{2m_\sigma\mu_\sigma(T)}
\label{pf0}
\end{equation}
are different between the two components, which is similar to an electron gas under an external magnetic field, as well as a spin-imbalanced Fermi gas. In that sense, the temperature $T$ works like an effective magnetic field in the presence of mass imbalance.  In addition, since the Fermi degeneracy of each component is dominated by the {\it scaled} temperature $T/T_{\rm F}^\sigma$, the heavy fermions become closer to the classical regime than the light mass component. (Note that $T/T_{\rm F}^{\rm H}>T/T_{\rm F}^{\rm L}$ when $m_{\rm H}>m_{\rm L}$.) Thus, the pseudogap associated with strong pairing fluctuations are expected to be more remarkable in the light mass component for a given temperature, although both the components {\it equally} contribute to the formation of preformed pairs, which is the origin of the pseudogap in the present system.
\par
The similarity between a mass-imbalanced Fermi gas and a spin-imbalanced Fermi gas at finite temperatures also implies that theoretical studies of the former system also meet the same problem as that known in the latter\cite{Liu2006,Parish2007a,Kashimura2012}. That is, in a spin-imbalanced Fermi gas, the Gaussian fluctuation theory developed by Nozi\`eres and Schmitt-Rink\cite{Nozieres1985,Melo1993,Ohashi2002}, as well as the (non-self-consistent) $T$-matrix approximation\cite{Perali2002}, that have been extensively used to successfully explain various BCS-BEC crossover physics in the absence of spin imbalance, breakdown\cite{Liu2006,Parish2007a,Kashimura2012}. Indeed, we later show that a mass-imbalanced Fermi gas also has the same problem. In the spin-imbalanced case, Ref.\cite{Kashimura2012} has recently presented a minimal extension of the $T$-matrix approximation to overcome this problem, by properly including higher order fluctuations beyond the $T$-matrix level. The calculated spin susceptibility in this extended $T$-matrix theory agrees well with the recent experiment on a $^6$Li Fermi gas\cite{Kashimura2012}. In this paper, we also employ this strategy, to assess the validity of the extended $T$-matrix approximation in the mass-imbalanced case.
\par
This paper is organized as follows. In Sec.II, we explain our extended $T$-matrix approximation (ETMA) for a two-component Fermi gas with mass imbalance. In Sec.III, we evaluate the superfluid phase transition temperature $T_{\rm c}$ in the BCS-BEC crossover region, to see how the ETMA gives improved results compared with the ordinary $T$-matrix approximation. In Sec.IV, we calculate single-particle excitations in the normal state above $T_{\rm c}$, to clarify effects of mass imbalance on the pseudogap phenomenon. Throughout this paper, we set $\hbar=k_{\rm B}=1$, and the system volume $V$ is taken to be unity. 
\par
\par
\section{Extended $T$-matrix theory of a mass-imbalanced Fermi gas}
\par
We consider a two-component uniform Fermi gas with mass imbalance, described by the Hamiltonian,
\begin{equation}
H = \sum_{\bm{p},\sigma}\xi_{\bm{p}\sig}c^{\dag}_{\bm p\sig}c_{\bm p\sig}
-U\sum_{\bm q}\sum_{\bm p, \bm{p}'}
c^{\dag}_{\bm p + \bm q/2,{\rm L}}c^\dag_{-\bm p + \bm q/2,{\rm H}}c_{-\bm {p}' + \bm q /2,{\rm H}}c_{\bm{p}'+\bm q /2,{\rm L}}.
\label{HAM}
\end{equation}
Here, $c_{\bm p\sig}$ is an annihilation operator of a Fermi atom with the kinetic energy $\xi_{\bm p\sig}=\varepsilon_{{\bm p}\sigma}-\mu_\sigma=p^2/(2m_\sig) - \mu_\sigma$, measured from the Fermi chemical potential $\mu_{\sigma}$, where  the labels $\sigma={\rm L},{\rm H}$ distinguish between the light mass ($m_{\rm L}$) component and heavy mass ($m_{\rm H}$) component. $-U$ $(<0)$ is a pairing interaction, which is assumed to be tunable by a Feshbach resonance\cite{Chin2010}. We actually measure the interaction strength in terms of the $s$-wave scattering length $a_s$, given by
\begin{equation}
\frac{4\pi a_s}{m}=\frac{-U}{1-U\sum_{\bm p}\frac{m}{p^2}},
\label{eq.as}
\end{equation}
where $m=2[m^{-1}_{{\rm L}}+m^{-1}_{\rm H}]^{-1}$ is twice the reduced mass. In this scale, the weak-coupling BCS regime and the strong-coupling BEC regime are characterized by $(k_{\rm F}a_s)^{-1}\lesssim -1$ and  $1 \lesssim (k_{\rm F}a_s)^{-1}$, respectively, where $k_{\rm F}=(3\pi N)^{1/3}$ is the Fermi momentum of a two-component gas of $N$ Fermi atoms (where $N$ is the total number of Fermi atoms). The region $-1 \lesssim (k_{\rm F}a_s)^{-1}\lesssim 1$ is called the BCS-BEC crossover region. 
\par
\begin{figure}
\begin{center}
\includegraphics[width=0.6\linewidth]{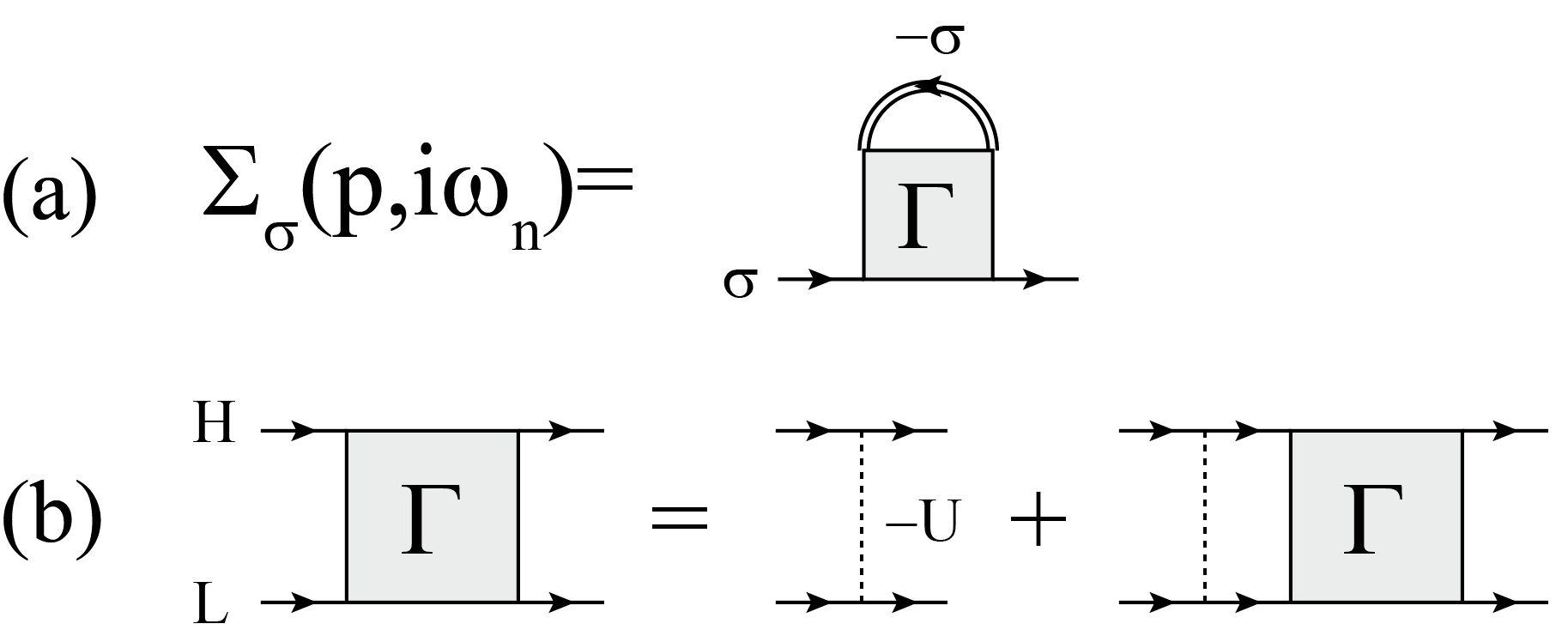}
\end{center}
\caption{(a) Self-energy correction $\Sigma_\sigma({\bm p}, i\omega_n)$ in the extended $T$-matrix approximation (ETMA). The double-solid line describes the dressed Green's function $G_\sigma$ in Eq. (\ref{etmag}). In this panel, $-\sigma$ means the opposite component to the $\sigma$-component. (b) Particle-particle scattering matrix $\Gamma({\bm q},i\nu_n)$. The solid line denotes the bare Green's function $G_\sigma^0$, and the dotted line describes the pairing interaction $-U$.}
\label{fig2}
\end{figure}
\par
Strong-coupling corrections to single-particle excitations are conveniently described by the self-energy $\Sigma_\sigma({\bm p},i\omega_n)$ in the single-particle thermal Green's function,
\begin{equation}
G_\sigma({\bm p},i\omega_n)=\frac{1}{i\om_n - \xi _{\bm p\sig}-\Sig_\sig(\bm p,i\om_n)},
\label{etmag}
\end{equation}
where $\omega_n$ is the fermion Matsubara frequency. In the extended $T$-matrix approximation (ETMA)\cite{Kashimura2012}, the self-energy is diagrammatically described as Fig. \ref{fig2}, which gives
\begin{equation}
\Sigma_\sigma({\bm p},i\om _n)=T\sum _{\bm q,\nu _n}\Gam(\bm q,i\nu _n)
G_{-\sig}({\bm q}-{\bm p},i\nu_n-i\om_n),
\label{etma}
\end{equation}
where $\nu_n$ is the boson Matsubara frequency. In Eq. (\ref{etma}), $-\sigma$ represents the opposite component to the $\sigma$-component. (For example, when $\sigma={\rm H}$, $-\sigma$ means the ${\rm L}$-component.) The particle-particle scattering matrix $\Gam(\bm q, i\nu _n)$ in Eq. (\ref{etma}) describes pairing fluctuations, having the form,
\begin{equation}
\Gam(\bm q, i\nu _n)=-{U  \over 1-U\Pi(\bm q,i\nu_n)},
\label{GAM}
\end{equation}
where
\begin{eqnarray}
\Pi(\bm q, i\nu_n)&=&T\sum_{\bm k,\om _n}G^0_{\rm L}(\bm k + \bm q/ 2,i\nu_n+i\om_n)G^0_{\rm H}(-\bm k+\bm q /2,-i\om_n)
\nonumber\\
&=&-\sum_{\bm k}\frac{1-f(\xi_{\bm{k}+\bm{q}/2,{\rm L}})-f(\xi_{-\bm{k}+\bm{q}/2,{\rm H}})}{i\nu_n-\xi_{\bm{k}+\bm{q}/2,{\rm L}}-\xi_{-\bm{k}+\bm{q}/2,{\rm H}}}
\label{PI}
\end{eqnarray}
is the lowest-order pair correlation function. In Eq. (\ref{PI}), $f(\varepsilon)=[e^{\varepsilon/T}+1]^{-1}$ is the Fermi distribution function. $G^0_\sigma({\bm p},i\om_n)=[i\om_n -\xi_{\bm p\sig}]^{-1}$ is the bare Green's function describing a free Fermi gas.
\par
The ordinary (non-self-consistent) $T$-matrix approximation (TMA) is obtained by simply replacing the dressed Green's function $G_{-\sigma}({\bm q}-{\bm p},i\nu_n-i\omega_n)$ in Eq. (\ref{etma}) with the {\it bare} one $G_{-\sigma}^0({\bm q}-{\bm p},i\nu_n-i\omega_n)$. Diagrammatically, it corresponds to the replacement of the double solid line ($G_{-\sigma}$) in Fig. \ref{fig2}(a) by the single solid line ($G_{-\sigma}^0$). 
\par 

The superfluid phase transition temperature $T_{\rm c}$ is determined from the Thouless criterion\cite{Thouless1960}, stating that the superfluid instability occurs when the particle-particle scattering matrix $\Gamma({\bm q},i\nu_n)$ has a pole at ${\bm q}=\nu_n=0$. The resulting equation for $T_{\rm c}$ is given by
\begin{eqnarray}
1&=&
UT\sum_{\bm k,\om _n}G^0_{\rm L}(\bm k,i\om_n)G^0_{\rm H}(-\bm k,-i\om_n)
\nonumber
\\
&=&
{U \over 2}\sum_{\bm k}
{\tanh(\xi_{{\bm k}{\rm L}}/(2T))+\tanh(\xi_{{\bm k}{\rm H}}/(2T)) 
\over \xi_{{\bm k}{\rm L}}+\xi_{{\bm k}{\rm H}}}.
\label{GAP}
\end{eqnarray}
We numerically solve the $T_{\rm c}$-equation (\ref{GAP}) under the assumption that $N_{\rm L}=N_{\rm H}=N/2$, where
\begin{eqnarray}
{N_\sigma}=T\sum_{\bm p,\om _n}G_{\sig}({\bm p},i\om_n)
\label{MMM}
\end{eqnarray}
is the number of Fermi atoms in the $\sigma$-component. For a given interaction strength, we self-consistently determine $T_{\rm c}$, $\mu_{\rm L}$, and $\mu_{\rm H}$ from the coupled equations (\ref{GAP}) and (\ref{MMM}). 
\par
Once $T_{\rm c}$ is determined, the chemical potentials $\mu_{\rm L}$ and $\mu_{\rm H}$ above $T_{\rm c}$ are simply determined from the number equation (\ref{MMM}). Then, the single-particle spectral weight $A_\sigma(\bm{p}, \omega)$, as well as the density of states $\rho_\sigma(\omega)$, in the normal state are evaluated from the analytic continued Green's function as\cite{Fetter1971},
\begin{eqnarray}
A_\sigma(\bm{p}, \omega) = -\frac{1}{\pi}{\rm Im}G_{\sigma}({\bm p},i\omega_n\rightarrow\omega+i\delta),
\label{spectral0}
\end{eqnarray}
\begin{eqnarray}
\rho_\sigma(\omega) = \sum_{\bm p}A_\sigma(\bm{p}, \omega),
\label{spectral}
\end{eqnarray}
where $\delta$ is an infinitesimally small positive number. 
\par
\par
\section{Superfluid phase transition temperature and effects of mass imbalance}
\par
\begin{figure}[t]
\begin{center}
\includegraphics[width=0.8\linewidth]{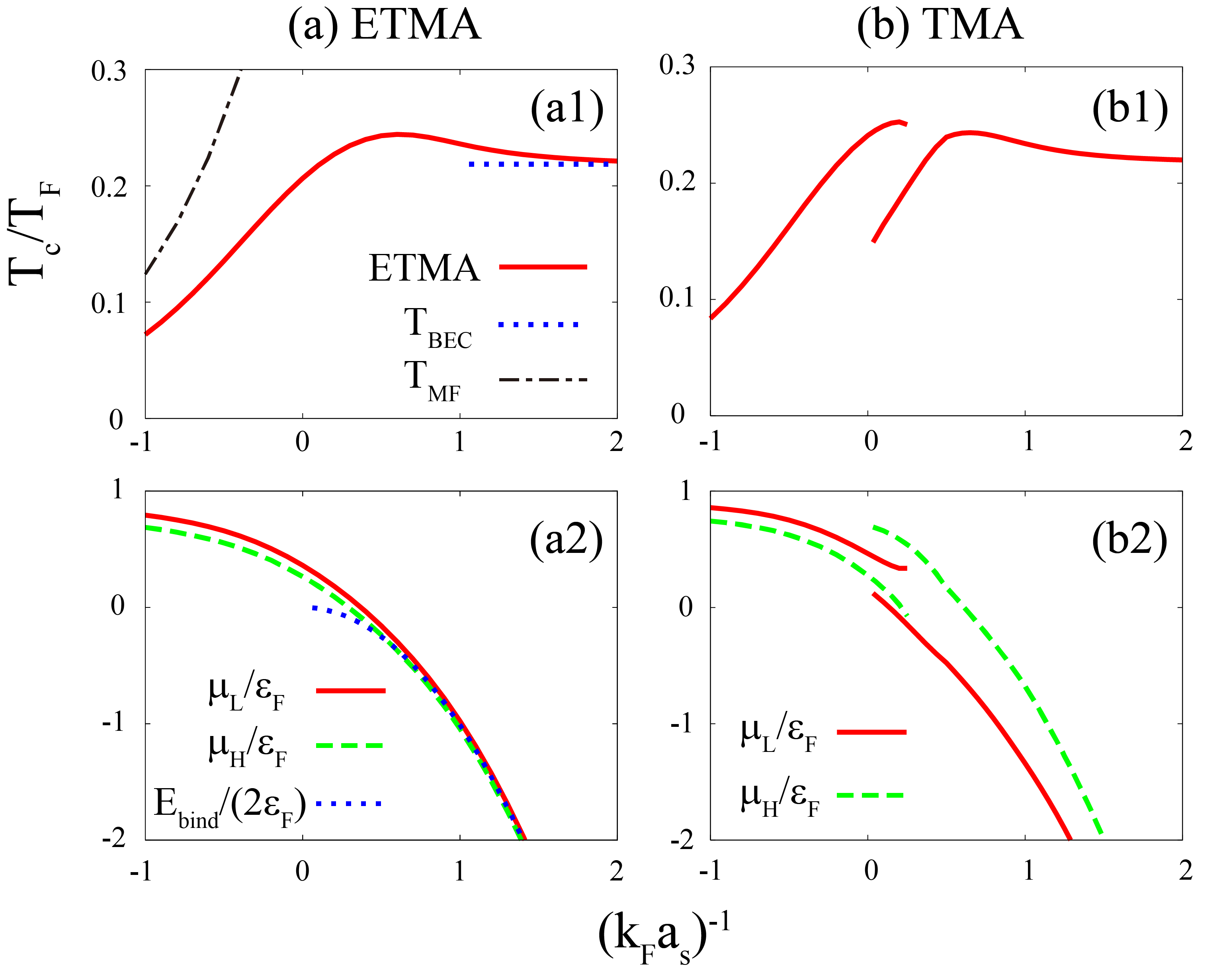}
\end{center}
\caption{(Color online) Self-consistent solutions of the coupled equations (\ref{GAP}) and (\ref{MMM}) when $m_{\rm L}/m_{\rm H}=0.9$. (a) Extended $T$-matrix approximation (ETMA). (b) $T$-matrix approximation (TMA). The upper panels show $T_{\rm c}$, and the lower panels show the Fermi chemical potential $\mu_\sigma$. $T_{\rm MF}$ is $T_{\rm c}$ in the weak-coupling BCS theory. $\varepsilon_{\rm F}~(=T_{\rm F})$ is the Fermi energy (Fermi temperature) of an assumed two-component gas of $N=N_{\rm L}+N_{\rm H}$ Fermi atoms with mass $m$.}
\label{fig3}
\end{figure}
\par

Figures \ref{fig3}(a1) and (a2) show the ETMA solutions of the coupled equations (\ref{GAP}) and (\ref{MMM}) when $m_{\rm L}/m_{\rm H}=0.9$. As in the mass-balanced case, $T_{\rm c}$ gradually deviates from the mean-field result ($T_{\rm MF}$) with increasing the interaction strength. In the BEC regime, $T_{\rm c}$ approaches a constant value, which is well described by the BEC phase transition temperature ($T_{\rm BEC}$) of a Bose gas of $N_{\rm B}=N/2$ molecules with mass $M=m_{\rm L}+m_{\rm H}$, given by
\begin{equation}
T_{\rm BEC}={2\pi \over M}\Bigl({N_{\rm B} \over \zeta(3/2)}\Bigr)^{2/3}.
\label{BEC}
\end{equation}
\par
The Fermi chemical potential $\mu_\sigma$ also exhibits the typical BCS-BEC crossover behavior, as shown in Fig. \ref{fig3}(a2). Starting from the weak-coupling regime, one sees that $\mu_{\rm \sigma}$ gradually decreases to be negative, as one passes through the BCS-BEC crossover region. In the BEC regime, $|\mu_{\rm L}+\mu_{\rm H}|$ approaches the binding energy of a two-body bound state, $E_{\rm bind}=1/(ma_s^2)$, as expected (where the effective mass $m$ is defined below Eq. (\ref{eq.as})).
\par 
In contrast, the ordinary $T$-matrix approximation (TMA) does not give the expected smooth BCS-BEC crossover in the presence of mass imbalance. In Figs. \ref{fig3}(b1) and (b2), we find that, although the ratio $m_{\rm L}/m_{\rm H}=0.9$ of the mass imbalance does not so deviate from unity in this figure, the TMA unphysically gives double-valued solutions around the unitarity limit.  We briefly note that a similar unphysical result is also obtained in the Gaussian fluctuation theory\cite{Nozieres1985} (although we do not explicitly show the results here). 
\par
\begin{figure}
\begin{center}
\includegraphics[width=0.4\linewidth]{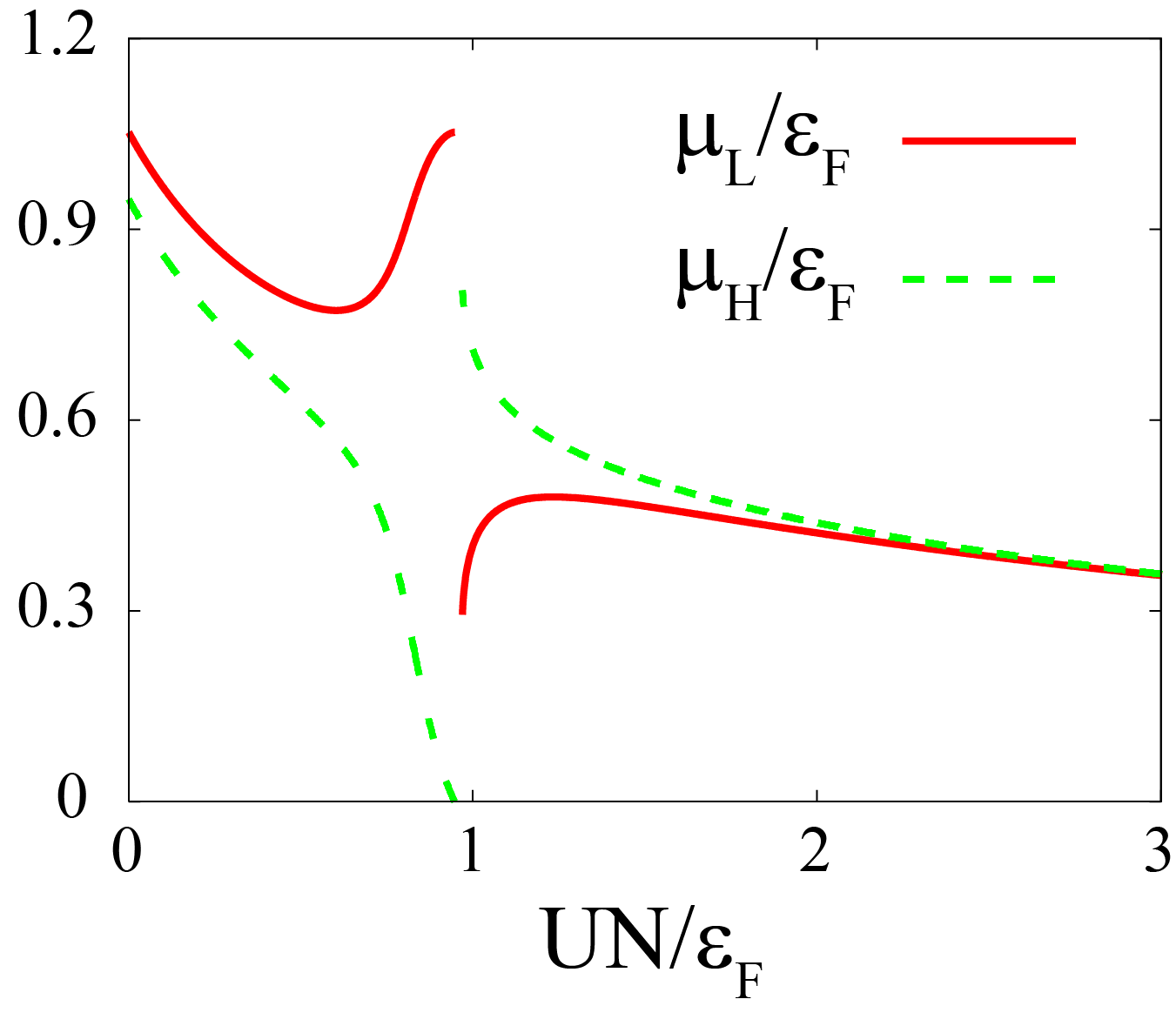}
\end{center}
\caption{(Color online) Calculated chemical potential $\mu_\sigma$, when the self-energy $\Sigma^{\rm TMA}_\sigma$ in Eq. (\ref{selfenergyTMA}) is used in the number equation (\ref{MMM}). We set $m_{\rm L}/m_{\rm H}=0.9$ and $T=0$.}
\label{fig4}
\end{figure}
\par
In a sense, the breakdown of the TMA shown in Fig. \ref{fig3}(b1) and (b2) has been already expected from the similarity between the mass-imbalanced system and the spin-imbalanced one.  As another viewpoint, however, we point out that the TMA involves an internal inconsistency in the presence of mass imbalance. To see this in a simple manner, we replace the particle-particle scattering matrix $\Gam(\bm q,i\nu _n)$ in Eq. (\ref{GAM}) by the bare interaction $-U$. In this case, the ETMA self-energy ($\Sigma_\sigma^{\rm ETMA}$) and the TMA self-energy ($\Sigma_\sigma^{\rm TMA}$) are simplified as, respectively
\begin{eqnarray}
\Sigma^{\rm ETMA}_\sigma = 
-UT\sum_{{\bm p},\om _n}G_{-\sigma}({\bm p},i\om_n)=-{UN \over 2}, 
\label{selfenergyETMA}
\end{eqnarray}
\begin{eqnarray}
\Sigma^{\rm TMA}_\sigma = 
-UT\sum_{{\bm p},\om _n}G^0_{-\sigma}({\bm p},i\om_n)=-UN_{-\sigma}^0, 
\label{selfenergyTMA}
\end{eqnarray}
where $N_\sigma^0 = T\sum_{{\bm p},\omega_n}G_\sigma^0({\bm p},i\omega_n)$. The ETMA self-energy in Eq. (\ref{selfenergyETMA}) reasonably describes the situation that a fermion in the $\sigma$-component interacts with $N/2$ fermions in the $-\sigma$-component. However, in the TMA case in Eq. (\ref{selfenergyTMA}), $N_{-\sigma}^0$ (which is evaluated by the bare Green's function) is usually different from $N/2$. Then, for example, in the extreme case with $\mu_{\rm L}>0$ and $\mu_{\rm H}=0$ at $T=0$ (which gives $N^0_{\rm L}=(2m_{\rm L}\mu_{\rm L})^{3/2}/(6\pi^2)>0$ and $N^0_{\rm H}=0$), Eq. (\ref{selfenergyTMA}) unphysically gives that, while a heavy fermion interacts with light fermions ($\Sigma^{\rm TMA}_{\rm H}\ne 0$), a light fermion behaves as a free particle ($\Sigma^{\rm TMA}_{\rm L}=0$), although the interaction occurs between the two components. Indeed, using Eq. (\ref{selfenergyTMA}) in the number equation (\ref{MMM}), we obtain a singular behaviors of $\mu_{\rm L}$ and $\mu_{\rm H}$ around $\mu_{\rm H}=0$, as shown in Fig. \ref{fig4}, which is somehow similar to the singularity seen in Fig. \ref{fig3}(b2).
\par
\begin{figure}
\begin{center}
\includegraphics[width=0.5\linewidth]{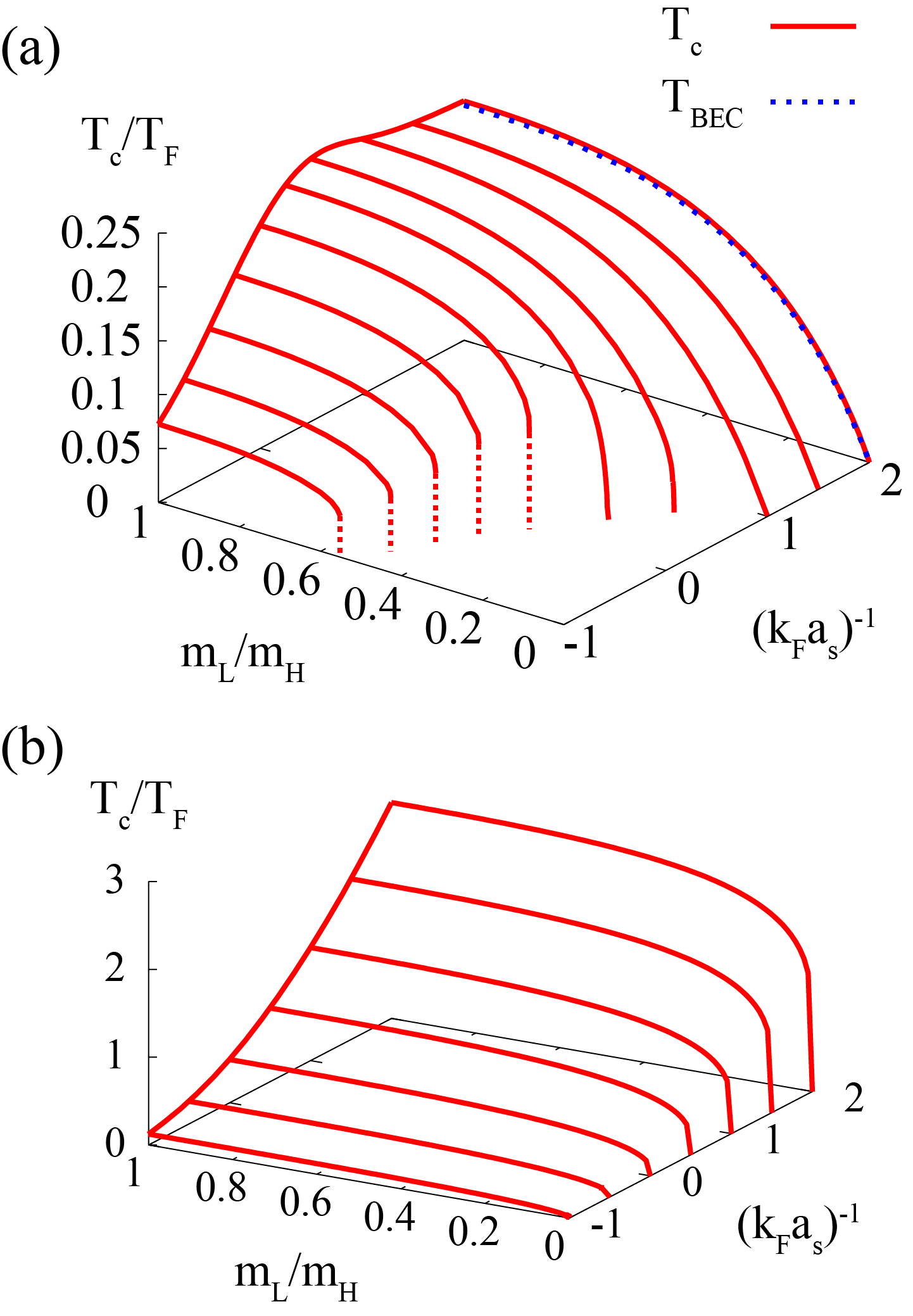}
\end{center}
\caption{(Color online) Calculated superfluid phase transition temperature $T_{\rm c}$ and effects of mass imbalance. (a) ETMA. (b) Mean-field theory. The dashed line shows $T_{\rm BEC}$ in Eq. (\ref{BEC}). In panel (a), $T_{\rm c}$ discontinuously vanishes in the BCS side, when the ratio $m_{\rm L}/m_{\rm H}$ becomes smaller than a critical value, which is shown as the dotted lines.}
\label{fig5}
\end{figure}
\par
Figure \ref{fig5}(a) shows the phase diagram of a mass-imbalanced Fermi gas obtained in the ETMA. The superfluid phase transition temperature $T_{\rm c}$ decreases, as the ratio $m_{\rm L}/m_{\rm H}$ decreases. In the BEC regime, the superfluid phase transition always occurs, irrespective of the value of $m_{\rm L}/m_{\rm H}$. Indeed, rewriting Eq. (\ref{BEC}) as,
\begin{equation}
T_{\rm BEC}={4m_{\rm L}/m_{\rm H} \over [1+(m_{\rm L}/m_{\rm H})]^2}
\times 0.218T_{\rm F},
\end{equation}
we find that $T_{\rm BEC}$ monotonically decreases to only vanish in the limit $m_{\rm L}/m_{\rm H}\to 0$.
\par
On the other hand, Fig. \ref{fig5}(a) shows that $T_{\rm c}$ vanishes at a finite value of $m_{\rm L}/m_{\rm H}~(>0)$ in the BCS regime. Rewriting the kinetic energies $\xi_{{\bm p}{\rm L}},\xi_{{\bm p}{\rm H}}$ in the forms
\begin{eqnarray}
\xi_{\bm p{\rm L}}&=&\frac{m}{m_{\rm L}}\left(\frac{p^2}{2m}-\mu \right ) - h, \label{xiup}\\
\xi_{\bm p{\rm H}}&=&\frac{m}{m_{\rm H}}\left(\frac{p^2}{2m}-\mu \right ) + h, \label{xidn}
\end{eqnarray}
(where $\mu=[\mu_{\rm L} + \mu_{\rm H}]/2$), one finds that the system is similar to that under an effective magnetic field, given by
\begin{eqnarray}
h={m_{\rm L}\mu_{\rm L}-m_{\rm H}\mu_{\rm H} \over m_{\rm L}+m_{\rm H}}.
\label{HHH}
\end{eqnarray}
Equation (\ref{HHH}) is usually finite when $m_{\rm L}\ne m_{\rm H}$, except for a free Fermi gas at $T=0$. Then, the mismatch of the Fermi surfaces between the two components occurs, leading to the vanishing $T_{\rm c}$ in the BCS regime.
\par
However, in {\it quantitatively} evaluating this pairing-breaking effect, one should note that the ETMA uses the bare Green's function $G_\sigma^0({\bm p},i\omega_n)$ in the particle-particle scattering matrix $\Gamma({\bm q},i\nu_n)$ in Eq. (\ref{GAM}). As a result, although the ETMA number equation (\ref{MMM}) includes many-body corrections through the self-energy $\Sigma_\sigma({\bm p},i\omega_n)$, the $T_{\rm c}$-equation (\ref{GAP}) is still at the mean-field level.
\par 
A problem coming from this inconsistency is that the above mentioned mismatch of the Fermi surfaces may be overestimated in the $T_{\rm c}$-equation. To see this in the BCS regime, we conveniently introduce the effective radius ${\bar k}_{{\rm F}\sigma}$ of the Fermi sphere in the $\sigma$-component from the pole of the analytic continued dressed Green's function at $\omega=0$ as
\begin{eqnarray}
{\bar k}_{\rm F\sigma}
&=&
\sqrt{{\tilde k}_{{\rm F}\sigma}^2+2m_\sigma\Sigma_\sigma({\bar k}_{{\rm F}\sigma},i\omega_n\to 0+i\delta)]}
\nonumber
\\
&\simeq&
{\tilde k}_{{\rm F}\sigma}
+{m_\sigma \over {\tilde k}_{{\rm F}\sigma} }
\Sigma_\sigma({\tilde k}_{{\rm F}\sigma},i\omega_n\to 0+i\delta),
\label{eq.radius}
\end{eqnarray}
where ${\tilde k}_{{\rm F}\sigma}=\sqrt{2m_\sigma\mu_\sigma}$. In obtaining the last expression, we have assumed a small self-energy. At very low temperatures, the Fermi surface sizes of the light mass and heavy mass components should be almost the same (${\bar k}_{{\rm F}\sigma={\rm L}}\simeq{\bar k}_{{\rm F}\sigma={\rm H}}$) in order to satisfy $N_{\rm L}=N_{\rm H}=N/2$. 
(We will confirm this from the analyses on the spectral weight $A_\sigma({\bm p},\omega)$ in Sec.IV.) In this case, noting that the second term in the last line in Eq. (\ref{eq.radius}) depends on $\sigma$, one expects that ${\tilde k}_{{\rm F},\sigma={\rm L}}\ne{\tilde k}_{{\rm F},\sigma={\rm H}}$. Since ${\tilde k}_{{\rm F}\sigma}=\sqrt{2m_\sigma\mu_\sigma}$ gives the effective Fermi surface size in the {\it bare} Green's function, the Fermi surface mismatch occurs in this Green's function, in spite that such a magnetic field effect is weak in the dressed Green's function. Since the ETMA $T_{\rm c}$-equation (\ref{GAP}) uses the bare Green's function, this inconsistency is considered to overestimate the suppression of $T_{\rm c}$ in the BCS regime.
\par
The above discussion indicates the importance of the consistent treatment of the number equation and the $T_{\rm c}$-equation in quantitatively evaluating $T_{\rm c}$ in a mass-imbalanced Fermi gas. In this sense, the mean-field BCS theory consisting of the $T_{\rm c}$-equation (\ref{GAP}) and the mean-field number equation,
\begin{equation}
N_\sigma=T\sum_{{\bm p},\omega_n}G_\sigma^0({\bm p},i\omega_n),
\label{MMM0}
\end{equation}
is consistent, because both the equations use the bare Green's function $G_\sigma^0({\bm p},i\omega_n)$. In this theory, we always obtain a finite $T_{\rm c}$ irrespective of the value of $m_{\rm L}/m_{\rm H}$, as shown in Fig. \ref{fig5}(b). Although the mean-field theory cannot describe the BCS-BEC crossover behavior of $T_{\rm c}$, it would be valid for the weak-coupling regime, so that this result also implies the overestimation of the effective magnetic field effect in the ETMA. To improve this point, it would be effective to employ the self-consistent $T$-matrix theory\cite{Haussmann}, where the dressed Green's functions are also used in the $T_{\rm c}$-equation. We will discuss this in our future paper\cite{Hanai}. In this paper, we examine single-particle properties of a mass-imbalanced Fermi gas within the framework of the ETMA.
\par
\begin{figure}[t]
\begin{center}
\includegraphics[width=0.8\linewidth]{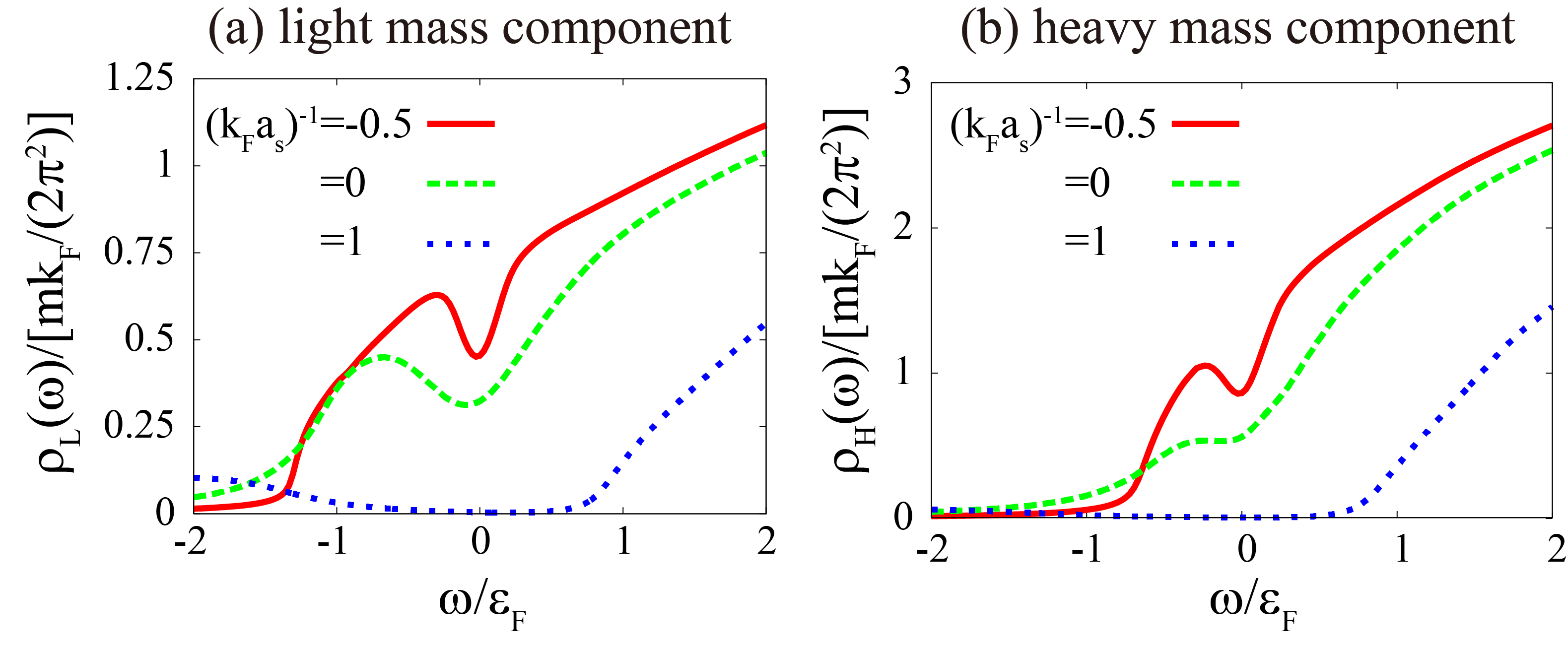}
\end{center}
\caption{(Color online) Calculated density of states $\rho_\sigma(\omega)$ at $T_{\rm c}$, when $m_{\rm L}/m_{\rm H}=0.5$. (a) Light mass component. (b) Heavy mass component.}
\label{fig6}
\end{figure}

\begin{figure}[t]
\begin{center}
\includegraphics[width=0.4\linewidth]{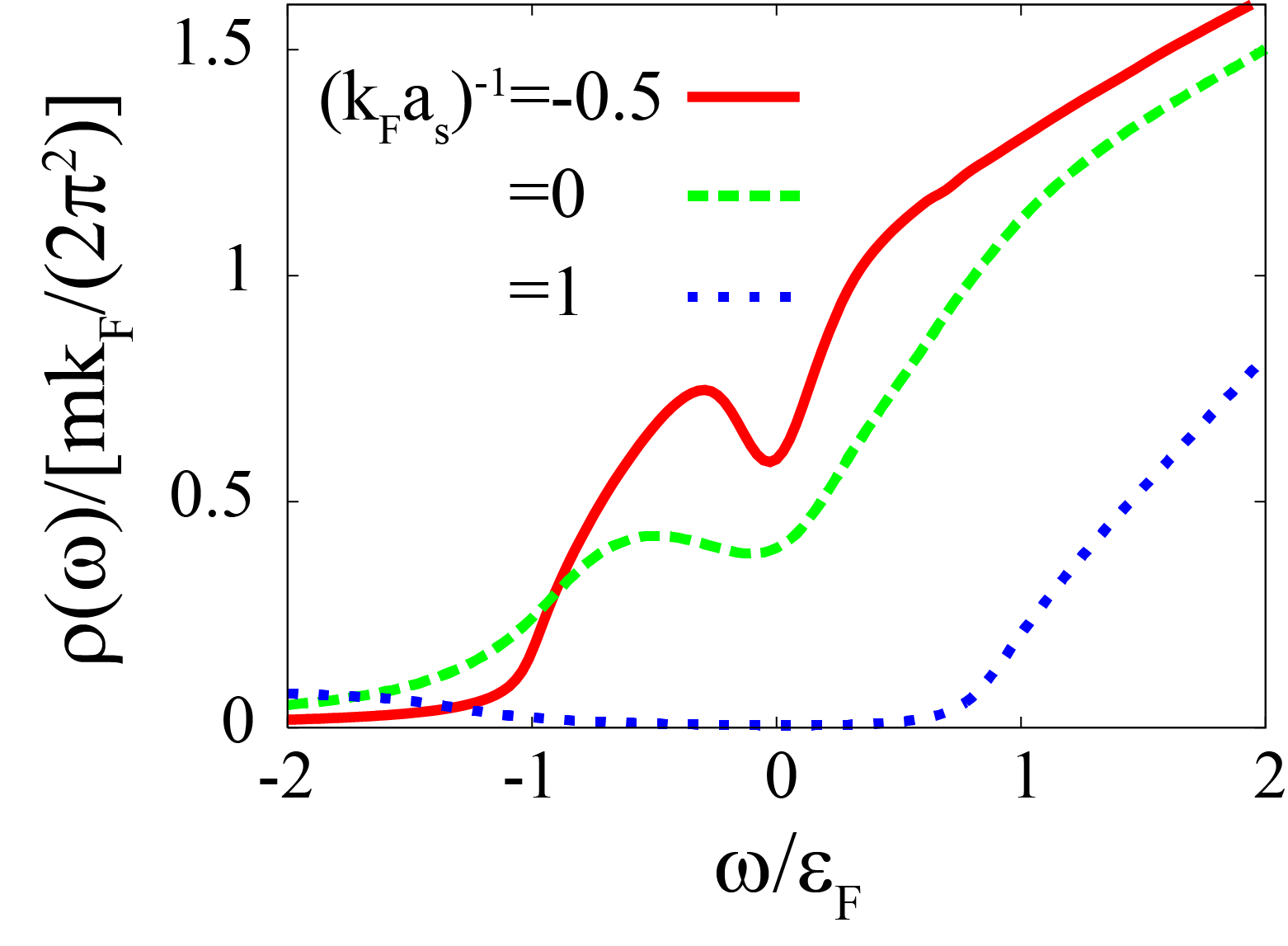}
\end{center}
\caption{(Color online) Calculated density of states $\rho(\omega)$ in the absence of mass imbalance ($m_{\rm L}/m_{\rm H}=1$). 
}
\label{fig7}
\end{figure}
\par
\section{Single-particle excitations in a mass-imbalanced Fermi gas}
\par
Figure \ref{fig6} shows the single-particle density of states $\rho_\sigma(\omega)$ in the BCS-BEC crossover regime of a mass-imbalanced Fermi gas at $T_{\rm c}$. In the BCS side at $(k_{\rm F}a_s)^{-1}=-0.5$, one sees dip structures in both the light and heavy mass components. Since the superfluid order parameter vanishes at $T_{\rm c}$, these are pseudogaps originating from pairing fluctuations. The pseudogap structures become more remarkable for a stronger pairing interaction. In the BEC regime when $(k_{\rm F}a_s)^{-1}=+1$, both $\rho_{\rm L}(\omega)$ and $\rho_{\rm H}(\omega)$ exhibit finite gaps ($\sim\varepsilon_{\rm F}$) associated with the binding energy of a two-body hetero-molecule. 
\par
\begin{figure}[t]
\begin{center}
\includegraphics[width=1\linewidth]{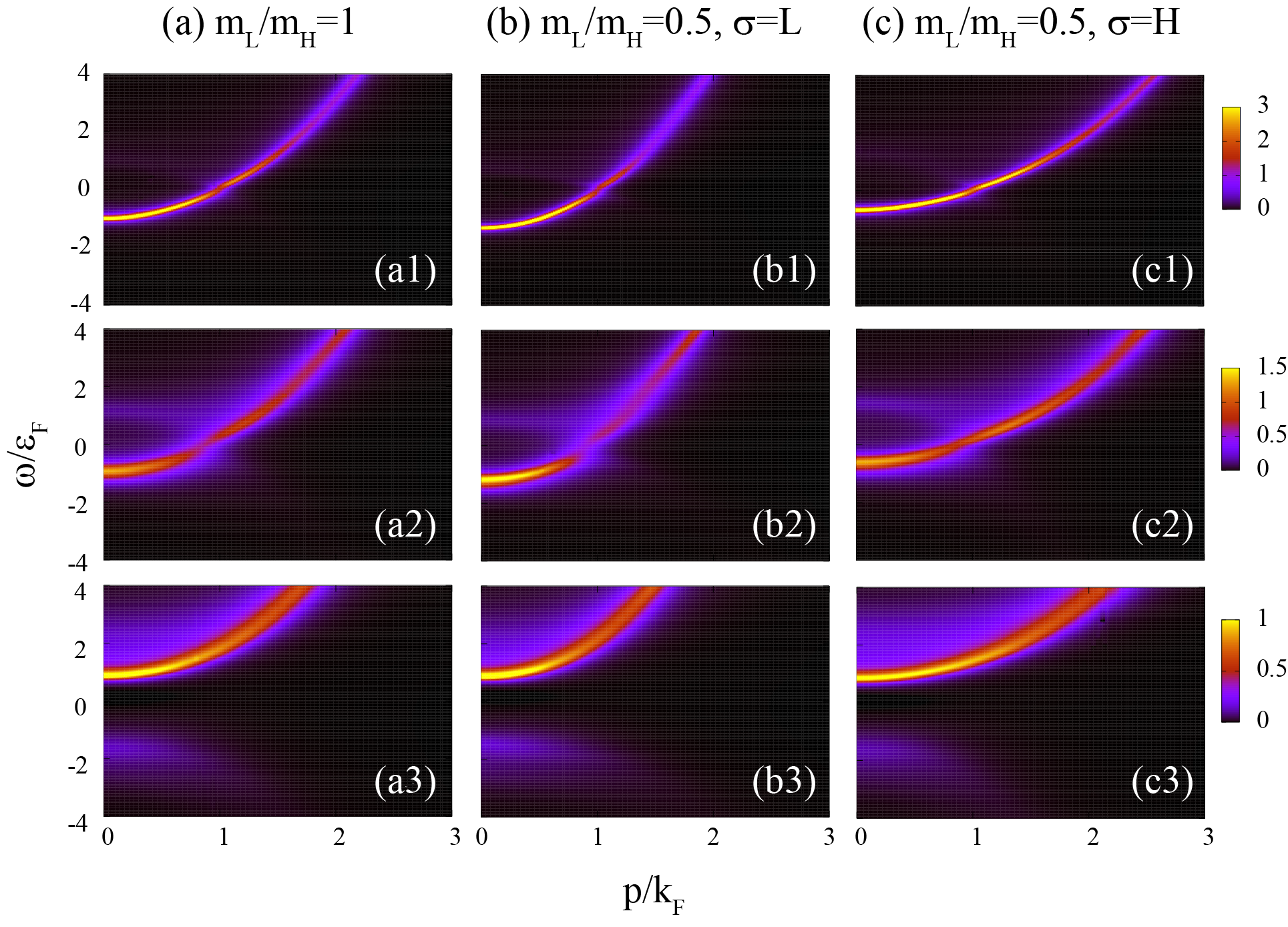}
\end{center}
\caption{(Color online) Calculated intensity of single-particle spectral weight $A_\sigma(\bm{p},\omega)$ at $T_{\rm c}$. Panels (a1)-(a3) show the mass-balanced case. Panels (b1)-(b3) (light mass component) and (c1)-(c3) (heavy mass component) show the mass-imbalanced case when $m_{\rm L}/m_{\rm H}=0.5$. We take the interaction strengths as $(k_{\rm F}a_s)^{-1}=-0.5$ (uppermost panels), $(k_{\rm F}a_s)^{-1}=0$ (middle panels), and $(k_{\rm F}a_s)^{-1}=1$ (lowest panels). The spectral intensity is scaled by $\varepsilon_{\rm F}^{-1}=2m/k_{\rm F}^2$. This normalization is also done in Figs. \ref{fig9}, \ref{fig12}, and \ref{fig14}.
}
\label{fig8}
\end{figure}
\par
Figure \ref{fig7} shows the density of states in the mass-balanced case. From the comparison of this figure with Fig. \ref{fig7}, pseudogap effects on the density of states are very similar between the mass-imbalanced case and the mass-balanced case. 
\par
This similarity at $T_{\rm c}$ is also seen in the single-particle spectral weight $A_\sigma({\bf p},\omega)$. In the mass-balanced case shown in Figs. \ref{fig8}(a1)-(a3), the spectral peak line gradually deviates from the free particle dispersion with increasing the interaction strength (panels (a1) and (a2)). In the unitarity limit (panel (a2)), the spectral peak around $\omega=0$ is affected by the pseudogap effect\cite{Tsuchiya2009b,Tsuchiya2010,Watanabe2010}. In the BEC regime (panel (a3)), the spectral weight exhibits a double peak structure, which just corresponds to the gap in Fig. \ref{fig7} at this interaction strength. These typical pseudogap behaviors in the BCS-BEC crossover region\cite{Chen2009,Magierski2009a,Tsuchiya2009b,Tsuchiya2010,Watanabe2010,Hu2010a,Stewart2008,Gaebler2010} are also seen in both the light mass component and the heavy mass component of a mass-balanced Fermi gas, as shown in Figs. \ref{fig8}(b1)-(b3) and (c1)-(c3), respectively.
\par
Since the dressed Green's function $G_\sigma$ is used in $A_\sigma({\bf p},\omega)$ (See Eq. (\ref{spectral0}).), the effective radius ${\bar k}_{{\rm F}\sigma}$ of the Fermi sphere in Eq. (\ref{eq.radius}) is directly related to the momentum at which the peak line in $A_\sigma({\bm p},\omega)$ crosses the zero-energy line ($\omega=0$). Although the precise determination of this quantity from $A_\sigma({\bm p},\omega)$ is actually not easy because of the pseudogap effect around $\omega=0$, we can still roughly estimate ${\bar k}_{{\rm F}\sigma}$ in the unitarity limit from Figs. \ref{fig8}(b2) and (c2) as
\begin{equation}
{\bar k}_{{\rm F}\sigma={\rm L}}
\simeq
{\bar k}_{{\rm F}\sigma={\rm H}}
\simeq k_{\rm F}.
\label{eq.300}
\end{equation}
Thus, as far as we treat the dressed Green's function $G_\sigma$ in Eq. (\ref{etmag}), the Fermi surface mismatch is almost absent even in the unitarity limit at $T_{\rm c}$. On the other hand, evaluating ${\tilde k}_{{\rm F}\sigma}$ in Eq. (\ref{pf0}), we obtain
\begin{equation}
{\tilde k}_{{\rm F}\sigma_{\rm L}}=\sqrt{2m_{\rm L}\mu_{\rm L}}=0.68p_{\rm F}
\gg
{\tilde k}_{{\rm F}\sigma_{\rm H}}=\sqrt{2m_{\rm H}\mu_{\rm H}}=0.14p_{\rm F}.
\end{equation}
As mentioned previously, this Fermi surface mismatch in the bare Green's function $G_\sigma^0$ directly affects the ETMA $T_{\rm c}$-equation (\ref{GAP}), leading to the suppression of $T_{\rm c}$. When the dressed Green's function $G_\sigma$ is also used in the $T_{\rm c}$-equation, such a depairing effect would be much weaker.
\par
In the ordinary $T$-matrix approximation, it has been shown that the pseudogap effect can be understood as a particle-hole coupling effect induced by pairing fluctuations\cite{Tsuchiya2009b}. To see how this picture is obtained in the present ETMA, it is convenient to approximate the self-energy $\Sigma_\sigma({\bm p},i\omega_n)$ in Eq. (\ref{etma}) to 
\begin{eqnarray}
\Sigma_\sigma(\bm{p},i\omega_n)
\simeq - G_{-\sigma}(-\bm{p},-i\omega_n )\Delta_{\rm pg}^2
=\frac{\Delta_{\rm pg}^2}{i\omega_n + \xi_{-p,-\sigma}+\Sigma_{-\sigma}(-\bm{p},-i\omega_n)}.
\label{eq.app_etma}
\end{eqnarray}
Here, $\Delta_{\rm pg}^2\equiv -T\sum_{\bm{q},\nu_n}\Gamma(\bm{q},i\nu_n)$ is the so-called pseudogap parameter\cite{Chen2009,Tsuchiya2009b}. In Eq. (\ref{eq.app_etma}), we have used the fact that pairing fluctuations described by $\Gamma({\bm q},i\nu_n)$ are enhanced in the low momentum and low energy region near $T_{\rm c}$. Equation (\ref{eq.app_etma}) gives
\begin{eqnarray}
\Sigma_\sigma(\bm{p},i\omega_n)
=
{{\tilde \Delta}_{{\rm pg}\sigma}^2(i\omega_n)
\over 
i\omega_n+\xi_{{\bm p},-\sigma}
},
\label{eq.ETMAappSigma}
\end{eqnarray}
where we have introduced the ETMA pseudogap parameter,
\begin{eqnarray}
{\tilde \Delta}_{{\rm pg}\sigma}^2(i\omega_n)=
{2\Delta_{\rm pg}^2 
\over 
1+\sqrt{1-
{\displaystyle 4\Delta^2_{\rm pg} 
\over \displaystyle
(i\omega_n-\xi_{{\bm p},\sigma})(i\omega_n+\xi_{{\bm p},-\sigma})
}}}.
\label{eq.ETMAapp}
\end{eqnarray}
Substituting Eq. (\ref{eq.ETMAappSigma}) into Eq. (\ref{etmag}), we obtain the BCS-type Green's function,
\begin{equation}
G_\sigma({\bm p},i\omega_n)=
{1 \over 
i\omega_n-\xi_{{\bm p},\sigma}-
{\displaystyle
{\tilde \Delta}_{{\rm pg}\sigma}^2(i\omega_n)
\over\displaystyle i\omega_n+\xi_{{\bm p},-\sigma}}},
\label{eq.ETMA2}
\end{equation}
which just exhibits the expected coupling between the particle branch ($\omega=\xi_{{\bm p},\sigma}$) and the hole branch ($\omega=-\xi_{{\bm p},-\sigma}$) by the ETMA pseudogap parameter ${\tilde \Delta}_{{\rm pg}\sigma}^2(i\omega_n)$. We briefly note that Eq. (\ref{eq.ETMA2}) is reduced to the previous result in the $T$-matrix approximation\cite{Tsuchiya2009b}, when one retains the term to $O(\Delta_{\rm pg}^2)$ in Eq. (\ref{eq.ETMAapp}).
\par
\begin{figure}[t]
\begin{center}
\includegraphics[width=0.7\linewidth]{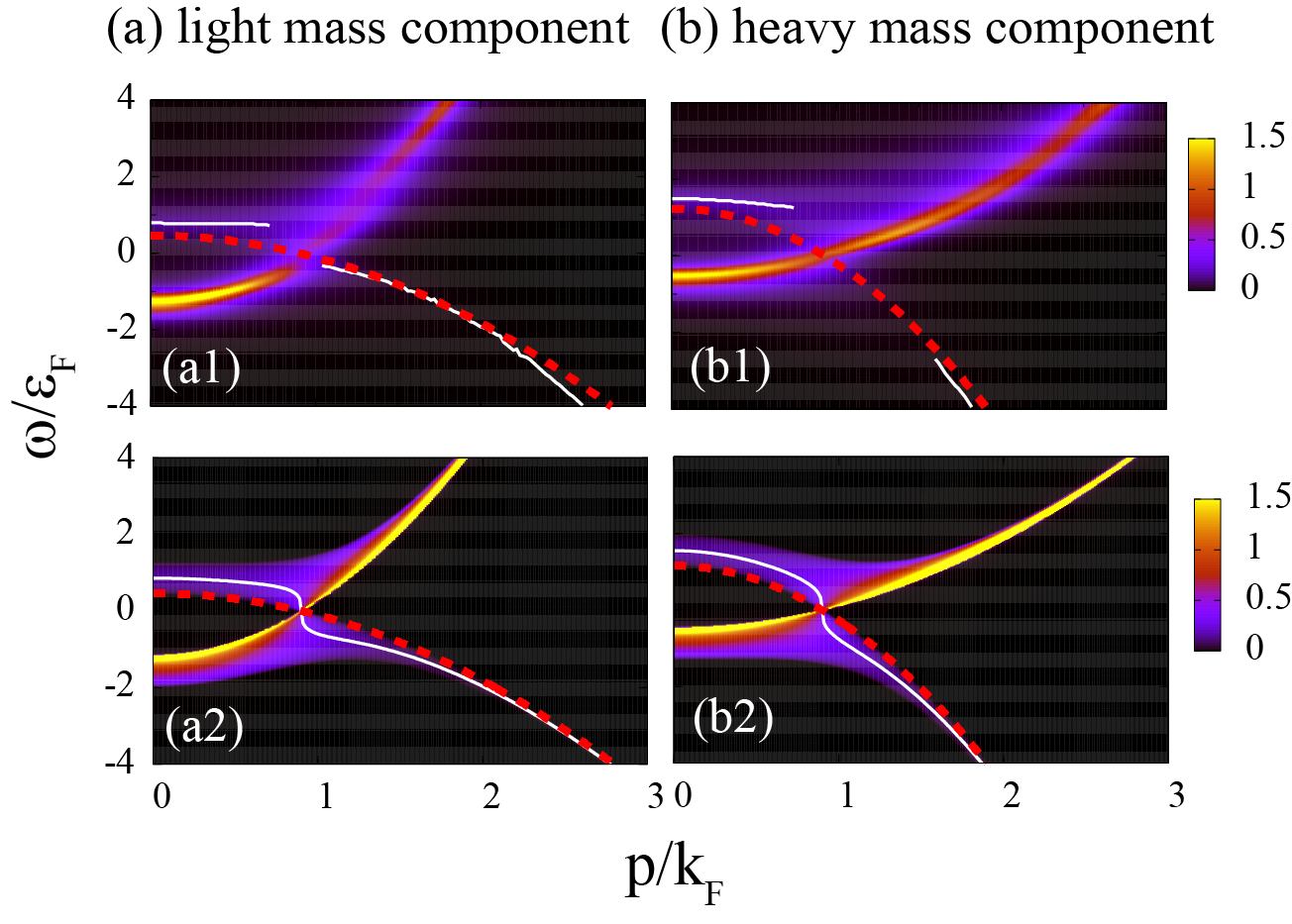}
\end{center}
\caption{(Color online) Comparison of the single-particle spectral weight $A_\sigma(\bm{p},\omega)$ in the unitarity limit at $T_{\rm c}$ when $m_{\rm L}/m_{\rm H}=0.4$. The upper panels are obtained in the ETMA, and the lower ones show the results when the static approximation in Eq. (\ref{eq.ETMA2}) is used. The light and left panels show $\sigma={\rm L}$ and $\sigma={\rm H}$, respectively. In each panel, the solid line shows the spectral peak, corresponding to the hole branch. For eye-guides, we also plot the hole dispersions $\omega=-[p^2/(2m_{\rm H})-{\bar k}_{\rm F,H}^2/(2m_{\rm H})]$ and $\omega=-[p^2/(2m_{\rm L})-{\bar k}_{\rm F,L}^2/(2m_{\rm H})]$ with ${\bar k}_{\rm F,L}={\bar k}_{\rm F,H}=0.9k_{\rm F}$ in panels (a1) and (b1), respectively (dashed lines). In obtaining the lower panels, we have taken $\Delta_{\rm pg}=0.7\varepsilon_{\rm F}$, and $\mu_{\rm \sigma}={\bar k}_{{\rm F},\sigma}^2/(2m_\sigma)$ so that the particle-hole coupling can occur at $p=0.9k_{\rm F}$.
}
\label{fig9}
\end{figure}
\par
However, since Eq. (\ref{eq.ETMA2}) only involves the coupling between the {\it bare} particle band ($\omega=\xi_{{\bm p},\sigma}$) and {\it bare} hole band ($\omega=-\xi_{{\bm p},-\sigma}$), this approximate treatment is not enough to {\it quantitatively} describe the pseudo-gapped spectral weight in the BCS-BEC crossover region. Indeed, in the unitarity limit of a mass imbalanced Fermi gas with $m_{\rm L}/m_{\rm H}=0.4$ at $T_{\rm c}$ (where the ETMA gives $\mu_{\rm L}=0.73\varepsilon_{\rm F}$ and $\mu_{\rm H}=-0.07\varepsilon_{\rm F}$), while Figs. \ref{fig9}(a1) and (b1) show that the particle-hole coupling occurs around $p=0.9k_{\rm F}$, the {\it bare} dispersion of the light mass component crosses the zero-energy line ($\omega=0$) at ${\tilde k}_{{\rm F},{\rm L}}=\sqrt{2m_{\rm L}\mu_{\rm L}}=0.5k_{\rm F}$, and that of the heavy mass component has no effective Fermi momentum ${\tilde k}_{{\rm F},{\rm H}}$ (because $\mu_{\rm H}=-0.07\varepsilon_{\rm F}<0$). Thus, when we use the ETMA results for $\mu_\sigma$ in Eq. (\ref{eq.ETMA2}), the particle-hole coupling at $p\simeq 0.9k_{\rm F}$ seen in the upper panels in Fig. \ref{fig9} is not reproduced. However, we briefly note that, when $\mu_{\rm L}$ and $\mu_{\rm H}$ in Eq. (\ref{eq.ETMA2}) are treated as fitting parameters, and they are chosen so that the single-particle dispersions $\omega=\xi_{{\bm p}\sigma}$ ($\sigma={\rm L,H}$) can cross the zero-energy line at $p\simeq 0.9k_{\rm F}$, the resulting spectral weight agrees well with the ETMA result when an appropriate value of the pseudogap parameter $\Delta_{\rm pg}$ is taken, as shown in Figs.\ref{fig9}(a2) and (b2). 
\par
\begin{figure}[t]
\begin{center}
\includegraphics[width=0.8\linewidth]{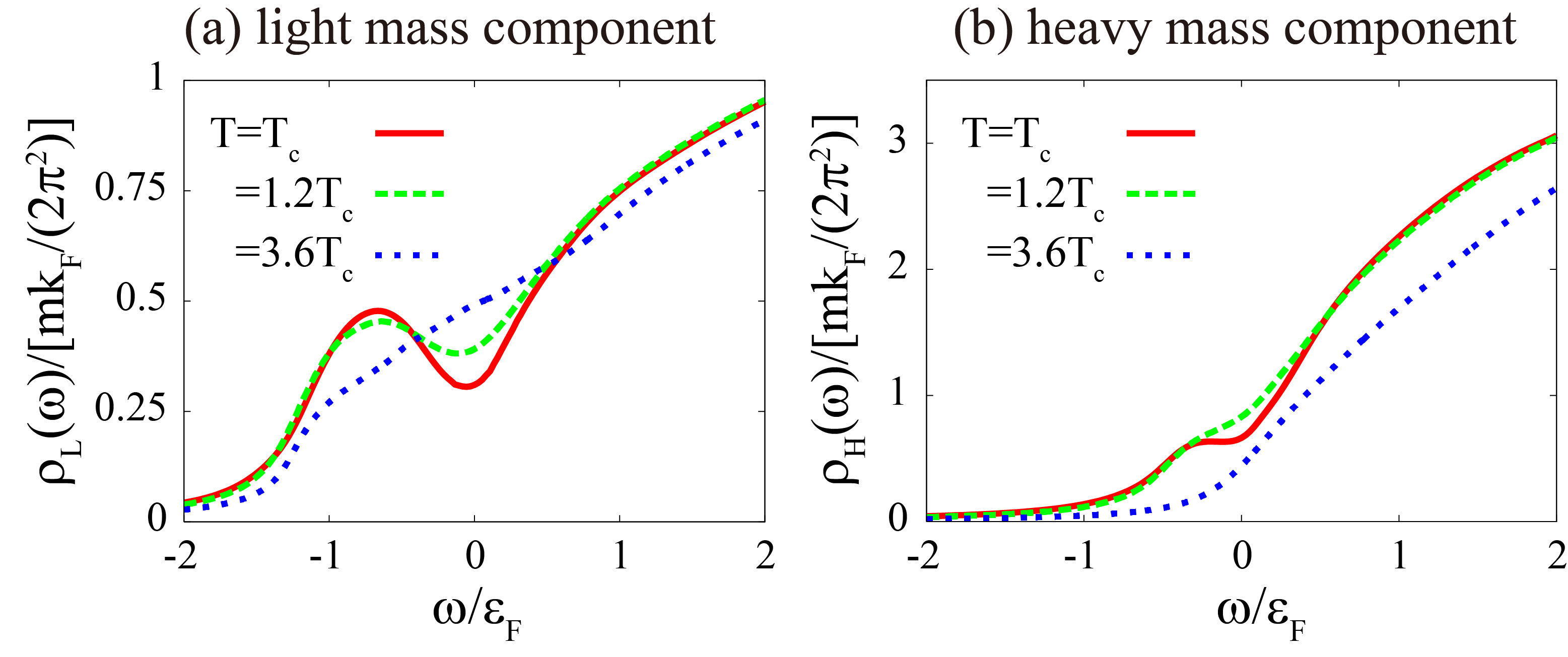}
\end{center}
\caption{(Color online) Calculated density of states $\rho_\sigma(\omega)$ at various temperatures above $T_{\rm c}$. (a) Light mass component. (b) Heavy mass component. We take $(k_{\rm F}a_s)^{-1}=0$, and $m_{\rm L}/m_{\rm H}=0.4$.
}
\label{fig10}
\end{figure}

\begin{figure}[t]
\begin{center}
\includegraphics[width=0.8\linewidth]{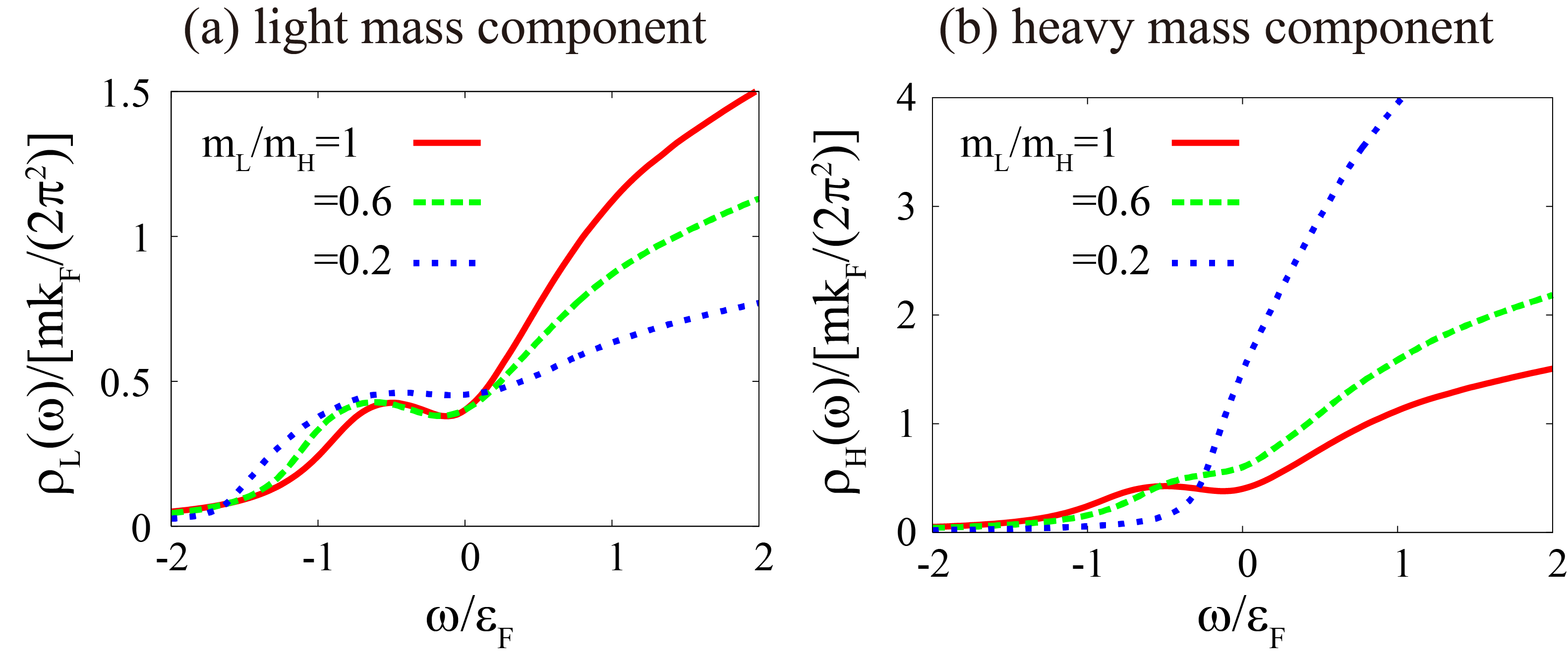}
\end{center}
\caption{(Color online) Calculated density of states $\rho_\sigma(\omega)$ at various ratios of mass imbalance $m_{\rm L}/m_{\rm H}$. (a) Light mass component. (b) Heavy mass component. We take $(k_{\rm F}a_s)^{-1}=0$, and $T=0.207T_{\rm F}$ (which equals $T_{\rm c}$ in the unitarity limit of a mass-balanced Fermi gas).}
\label{fig11}
\end{figure}

\begin{figure}[!h]
\begin{center}
\includegraphics[width=0.7\linewidth]{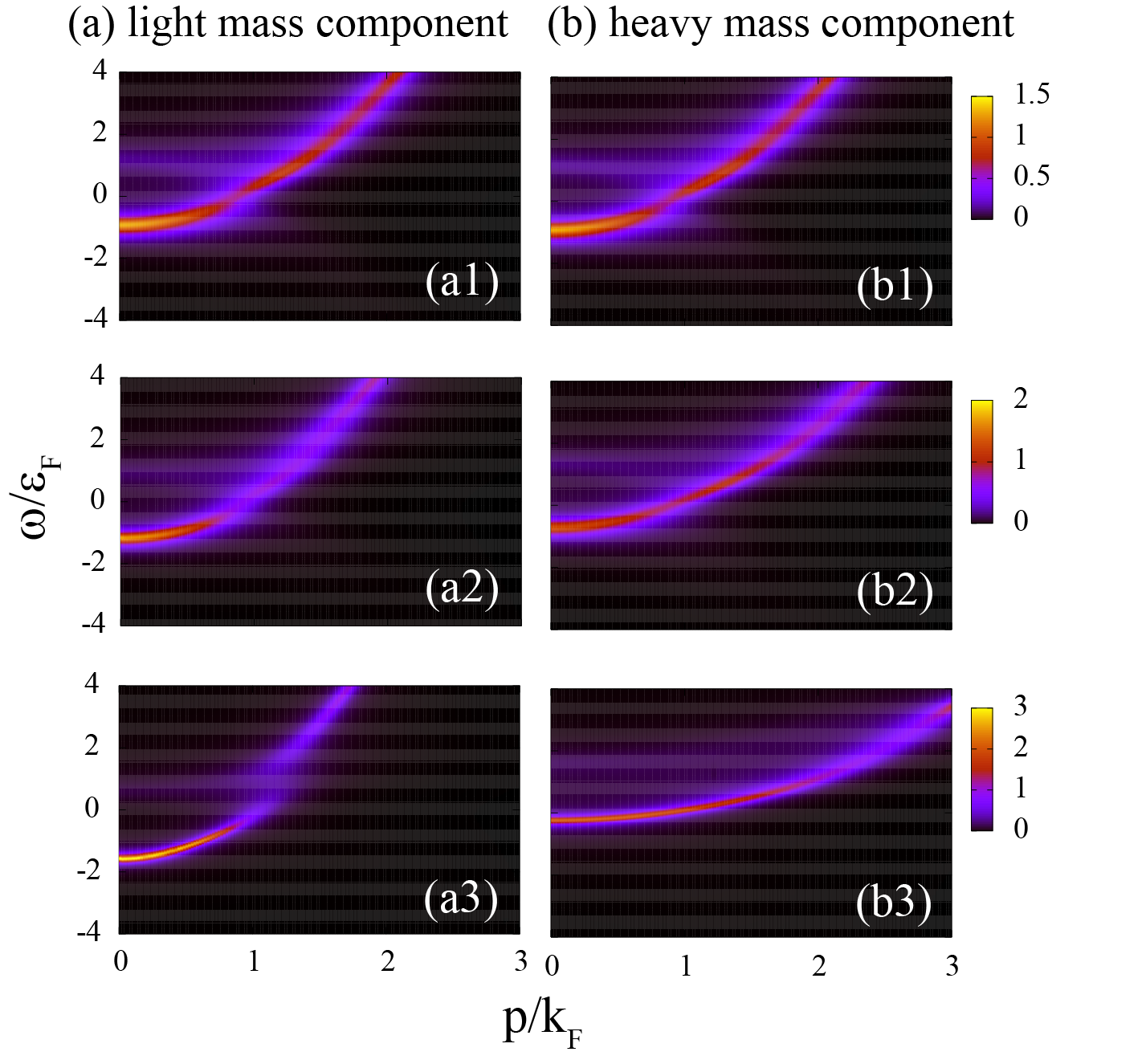}
\end{center}
\caption{(Color online) Calculated intensity of the spectral weight $A_\sigma(\bm{p},\omega)$ in the unitarity limit at $T=0.207T_{\rm F}$. The right and left panels show the light mass and heavy mass components, respectively. The ratio of the mass imbalance equals $m_{\rm L}/m_{\rm H}=1$ (upper most panels), 0.6 (middle panels), and 0.2 (lowest panels).
}
\label{fig12}
\end{figure}
\par
Now we consider the region above $T_{\rm c}$. Figure \ref{fig10} shows the single-particle density of states $\rho_\sigma(\omega)$ in the normal state. When we define the pseudogap temperature $T^*_\sigma$ as the temperature at which the pseudogap (dip) structure disappears in $\rho_\sigma(\omega)$, this figure indicates that $T_{\rm L}^*>T_{\rm H}^*\sim 1.2T_{\rm c}$. Thus, the pseudogap only appears in $\rho_{\rm L}(\omega)$ in the temperature region, $T_{\rm L}^*>T>T_{\rm H}^*$. Since the light and heavy fermions {\it equally} contribute to the formation of preformed pairs (which is the origin of the pseudogap), this result looks somehow strange. However, the key is that the Fermi temperatures are different between the two components as $T_{\rm F}^{\rm L}>T_{\rm F}^{\rm H}$. Because of this, the difference $[T-T_{\rm c}]/T_{\rm F}^\sigma$ of the temperature from $T_{\rm c}$ {\it scaled by} $T_{\rm F}^\sigma$ is larger in the heavy mass component than in the light mass component. That is, the former is effectively further away from the superfluid instability, leading to the lower pseudogap temperature $T_{\rm H}^*<T_{\rm L}^*$. Since the pseudogap phenomenon equally occurs in the two components in an ordinary mass-balanced Fermi gas, this is a characteristic phenomenon in the presence of mass imbalance.
\par
Such a component-dependent pseudogap phenomenon is also obtained, when one adjusts the ratio $m_{\rm L}/m_{\rm H}$ at a fixed temperature. In Fig. \ref{fig11}, while the pseudogap structure in $\rho_{\rm L}(\omega)$ remains even at $m_{\rm L}/m_{\rm H}=0.2$ (panel (a)), the dip structure no longer exists in $\rho_{\rm H}(\omega)$ at this ratio of mass imbalance (panel (b)). In this case, although the temperature is fixed, since $T_{\rm c}$ is lowered by mass imbalance (See Fig. \ref{fig5}(a).), we again obtain a larger scaled temperature difference $[T-T_{\rm c}(m_{\rm L}/m_{\rm H}<1)]/T_{\rm F}^\sigma$ in the heavy mass component\cite{NOTE3}, leading to weaker strong-coupling effects in this component. 
\par
As expected, Fig. \ref{fig12} shows that the component-dependent pseudogap phenomena also appears in the particle-hole coupling effect in $A_\sigma({\bm p},\omega)$. In the light mass component shown in panels (a1)-(a3), a hole branch, as well as the modification of a particle branch around $p=k_{\rm F}$ by the particle-hole coupling effect, remain down to $m_{\rm L}/m_{\rm H}=0.2$. On the other hand, the particle-hole coupling gradually disappears with decreasing the ratio $m_{\rm L}/m_{\rm H}$, and eventually the particle branch only remains at $m_{\rm L}/m_{\rm H}=0.2$ in the heavy mass component (panels (b1)-(b3)).
\par
\begin{figure}
\begin{center}
\includegraphics[width=0.8\linewidth]{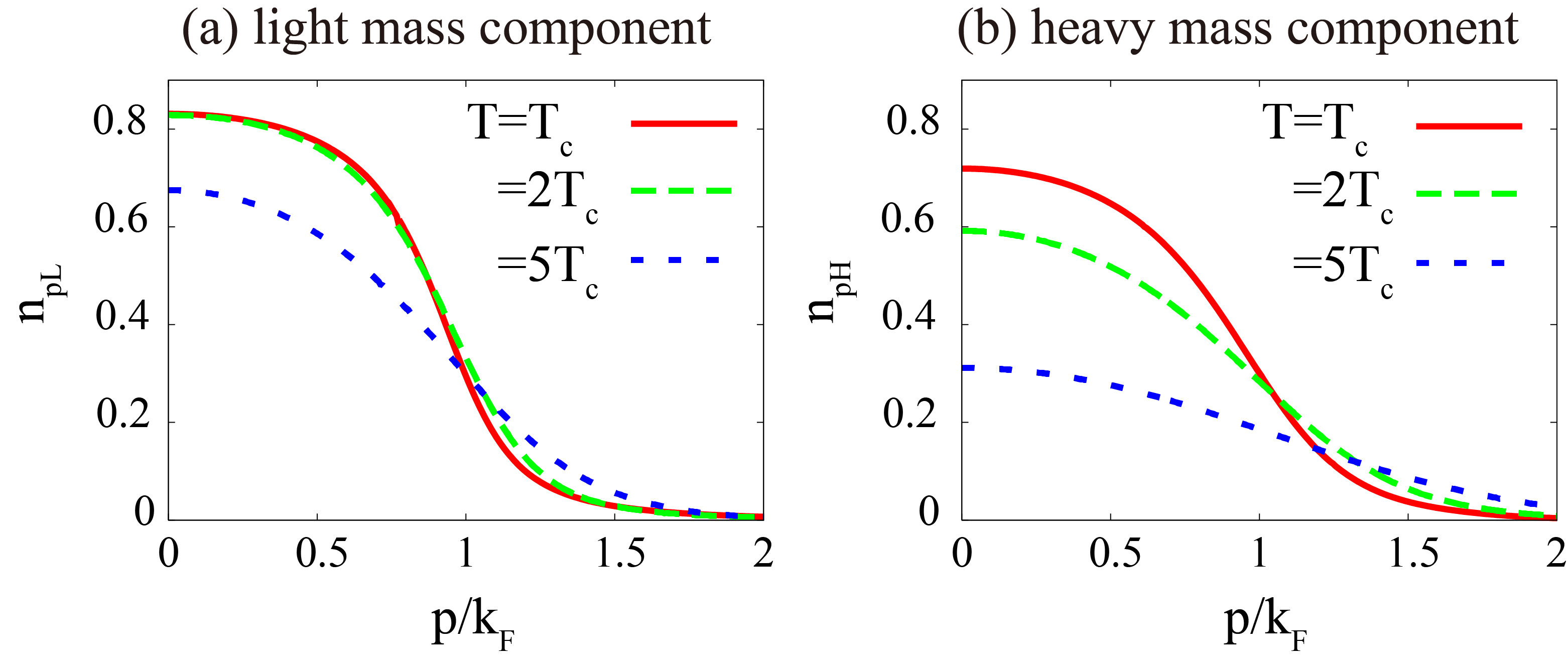}
\end{center}
\caption{(Color online) Momentum distribution $n_{{\bm p}\sigma}$ in a mass imbalanced Fermi gas. (a) Light mass component. (b) Heavy mass component. We take $(k_{\rm F}a_s)^{-1}=0$, and $m_{\rm L}/m_{\rm H}=0.4$.}
\label{fig13}
\end{figure}
\par
\begin{figure}[!h]
\begin{center}
\includegraphics[width=0.7\linewidth]{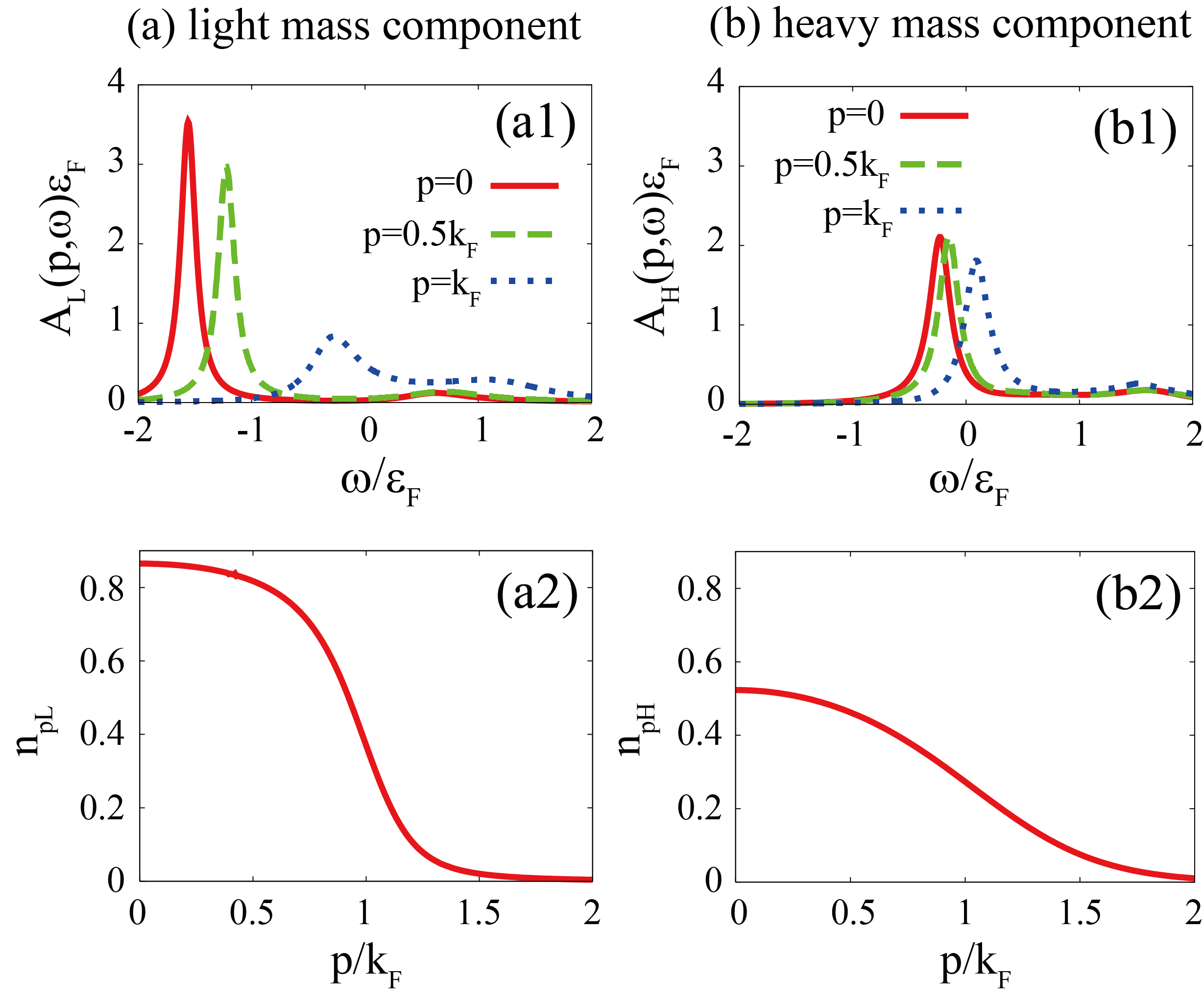}
\end{center}
\caption{(Color online) Calculated spectral weight $A_\sigma({\bm p},\omega)$ as a function of energy $\omega$ (upper panels). The lower panels show the the momentum distribution $n_{{\bm p}\sigma}$ in Eq. (\ref{np}). In this figure, the left and right panels show the results for the light and heavy mass components, respectively. We take $(k_{\rm F}a_s)^{-1}=0$, $m_{\rm L}/m_{\rm H}=0.2$, and $T=0.207T_{\rm F}$.
}
\label{fig14}
\end{figure}
\par
The different scaled temperatures between the light mass component and heavy mass component also affect their momentum distributions, given by
\begin{equation}
n_{{\bm p}\sigma}=
\langle c_{{\bm p}\sigma}^\dagger c_{{\bm p}\sigma}\rangle
=
T\sum_{\omega_n}G_\sigma({\bm p},i\omega_n).
\label{np}
\end{equation}
In Fig. \ref{fig13}, we clearly see that the broadening of the Fermi edge at $p=k_{\rm F}$ by thermal fluctuations are more remarkable in the heavy mass component, compared with the light mass component, which is simply due to the fact of $T/T_{\rm F}^{\rm H}>T/T_{\rm F}^{\rm L}$. In addition, we also find that $n_{{\bm p}{\rm H}}$ is more sensitive to the temperature than $n_{{\bm p}{\rm L}}$, because of the same reason.
\par
Figure \ref{fig14}(a1) shows the spectral weight $A_{\rm L}({\bm p},\omega)$ as a function of energy $\omega$. In the low momentum region ($p<k_{\rm F}$), since the occupation number $n_{{\bm p}{\rm L}}$ of light fermions is close to unity (See panel (a2).), the so-called Pauli blocking works to some extent there, which suppresses particle scatterings in this regime. As a result, the spectral peak in the low momentum region is sharp. (See the results at $p=0$ and $0.5k_{\rm F}$ in panel (a1).) On the other hand, scatterings of a light fermion frequently occur near the Fermi surface ($p\sim k_{\rm F}$), a broad spectral peak is obtained at $p=k_{\rm F}$ in Fig. \ref{fig14}(a1). 
\par
In contrast, the Fermi degeneracy in the region $p\lesssim k_{\rm F}$ is almost lifted in the heavy mass component, as shown in Fig. \ref{fig14}(b2). In addition, the momentum dependence of $n_{{\bm p}{\rm H}}$ shown in panel (b2) is weak, compared with the case of light mass component shown in panel (a2). Thus, while the heavy fermions in the low momentum region can be easily scattered, the momentum dependence of scattering effects is weak. As a result, the width of the spectral peak shown in Fig. \ref{fig14}(b1) is insensitive to the momentum $p$ in the region $0\le p\lesssim k_{\rm F}$, and the peak width at $p=0$ is broader than the case of light mass component shown in panel (a1).
\par
\section{Summary}
\par
To summarize, we have discussed single-particle properties of a mass-imbalanced Fermi gas in the BCS-BEC crossover region. The ordinary (non-self-consistent) $T$-matrix approximation, which is known to be a powerful strong-coupling theory to study various BCS-BEC crossover physics in the mass-balanced case, does not work in the presence of mass imbalance. We overcame this serious problem, by employing the extended $T$-matrix theory (ETMA) developed in the case of a spin-imbalanced Fermi gas. Using this, we determined the phase diagram of a Fermi gas in terms of the strength of a pairing interaction, as well as the ratio of mass imbalance. 
\par
Within the framework of the ETMA, we calculated the single-particle density of states $\rho_\sigma(\omega)$, as well as the spectral weight $A_{\sigma}({\bm p},\omega)$, to see how the pseudogap phenomenon in the BCS-BEC crossover region is affected by the presence of mass imbalance. At $T_{\rm c}$, the pseudogap appears in these quantities in both the light mass component and heavy mass component, which becomes more remarkable with the increase in the interaction strength. In this sense, the pseudogap phenomena in the presence of mass imbalance is very similar to the ordinary mass-balanced case. 
\par
Effects of mass imbalance on the pseudogap phenomenon become remarkable, as one raises the temperature from $T_{\rm c}$. Since the Fermi temperature $T_{\rm F}^{\rm H}$ in the heavy mass component is lower than that in the light mass component $T_{\rm F}^{\rm L}$, heavy fermions feel higher {\it scaled} temperature, $T/T_{\rm F}^{\rm H}>T/T_{\rm F}^{\rm L}$. Thus, the pseudogap in $\rho_\sigma(\omega)$ disappears at a lower temperature in the heavy mass component than in the light mass component. In addition, since the Pauli blocking inside the Fermi surface works more effectively in the light mass component because of the lower scaled temperature, the momentum dependence of the peak width in the single-particle spectral weight $A_\sigma({\bm p},\omega)$ is more remarkable in this component. That is, while a sharp spectral peak is obtained deep inside the Fermi level (because of the suppression of particle scatterings by the Pauli's exclusion principle), the spectral peak becomes broad near the Fermi surface (where particle scatterings can frequently occur). In contrast, since the heavy mass component is closer to the classical regime because of the higher scaled temperature, $T/T_{\rm F}^{\rm H}>T/T_{\rm F}^{\rm L}$, particle scatterings deep inside the Fermi surface are not so suppressed as in the case of light fermions. As a result, the peak width of the spectral weight in the heavy mass component becomes broad even around $p=0$, compared with that in the light mass component. In addition, the momentum dependence of the peak width in the heavy mass component becomes weaker. Since these different pseudogap phenomena between the two components never occur in the absence of mass imbalance, the observation of these component-dependent pseudogap phenomena by the photoemission-type experiment\cite{Stewart2008,Gaebler2010} would be an interesting challenge.
\par
We note that, although the extended $T$-matrix approximation (ETMA) can overcome the problem existing in the ordinary $T$-matrix approximation, it still has room for improvement. In particular, the ETMA is not a fully self-consistent approximation in the sense that the particle-particle scattering matrix uses the bare Green's function. This inconsistent treatment is considered to overestimate the pair-breaking effect associated with mass imbalance, especially in the weak-coupling BCS regime. Thus, although the present ETMA gives the vanishing $T_{\rm c}$ in the BCS regime when $m_{\rm L}/m_{\rm H}\ll 1$, we need further analyses on this regime beyond the ETMA. In this regard, an extension of the self-consistent $T$-matrix theory to a mass-imbalanced Fermi gas would be useful. We will discuss this extension in our future paper\cite{Hanai}.
\par
In this paper, we have simply treated a uniform system, ignoring effects of a harmonic trap. Since different species may feel different trap potentials, when we include effects of a harmonic trap, we need to consider the situation that the two components exhibit different density profile, leading to spatially dependent population imbalance. Thus, to examine a real trapped mass-imbalanced Fermi gas, inclusion of both the mass imbalance and spin imbalance would be important, which is also an exciting future problem. 
\par
So far, only the simplest isotropic $s$-wave superfluid state has been realized in ultracold Fermi gases. Although the study of this system is a crucial issue for the understanding of BCS-BEC crossover physics, the search for a more complicated pairing state would be also an important challenge for the further development of this research field. In addition, Fermi superfluids consisting of different species have recently discussed in various research fields, such as excitons in semiconductor physics, exciton-polaritons in semiconductor optics, and color superconductivity in high-energy physics. Thus, the realization of a mass-imbalanced superfluid Fermi gas would contribute to the study of these systems. Since the pseudogap phenomenon is a precursor of the superfluid instability, the observation of this would be helpful to see to what extent the system is close to the superfluid phase transition temperature. Thus, our results would be useful for, not only the understanding of a Fermi superfluid in the presence of mass imbalance, but also the realization of this hetero-pairing state.
\par
\begin{acknowledgements}
We thank S. Watabe for useful discussions. This work was supported by Institutional Program for Young Researcher Oversea Visits from the Japan Society for the Promotion of Science. Y. O. was supported by Grant-in-Aid for Scientific research from MEXT in Japan (22540412, 23104723, 23500056).
\end{acknowledgements}


\begin{thebibliography}{99}
\bibitem{NOTE3} We briefly note that the ETMA gives the vanishing $T_{\rm c}$ in the unitarity limit, when $m_{\rm L}/m_{\rm H}\le 0.32$.
\end{thebibliography}
\end{document}